\documentclass[twocolumn,preprintnumbers,superscriptaddress,amsmath,amssymb]{revtex4}

\usepackage{amsmath}
\usepackage{graphicx}
\usepackage{dcolumn}
\usepackage{bm}
\usepackage{color}
\usepackage{booktabs, longtable}
\usepackage{multirow}
\usepackage{CJK}
\usepackage{ltxtable}
\usepackage{rotating}
\usepackage[colorlinks, citecolor=red]{hyperref}
\usepackage{verbatim}
\usepackage{amsbsy}
\usepackage{amsfonts}
\usepackage{mathrsfs}
\usepackage[english]{babel}

\begin{document}

\title{Ground-state properties and structure evolutions of odd-$A$ transuranium Bk isotopes by deformed relativistic Hartree-Bogoliubov theory in continuum}

\author{Zi-Dan Huang}
\affiliation{School of Physics, Zhengzhou University, Zhengzhou 450001, China}

\author{Wei Zhang}
\affiliation{School of Physics, Zhengzhou University, Zhengzhou 450001, China}

\author{Shuang-Quan Zhang}
\affiliation{State Key Laboratory of Nuclear Physics and Technology, School of Physics, Peking University, Beijing 100871, China}

\author{Ting-Ting Sun}
\email[Corresponding author, ]{ttsunphy@zzu.edu.cn}
\affiliation{School of Physics, Zhengzhou University, Zhengzhou 450001, China}

\begin{abstract}
The studies of transuranium nuclei are of vital significance in exploring the existence of the ``island of superheavy nuclei". This work presents the systematic investigations for the ground-state properties and structure evolutions of odd-$A$ transuranium Bk isotopes taking the deformed relativistic Hartree-Bogoliubov theory in continuum~(DRHBc) with PC-PK1 density functional, in comparison with those by spherical relativistic continuum Hartree-Bogoliubov~(RCHB) theory. The DRHBc calculations offer improved descriptions of the binding energies, closely aligning with the experimental data. The incorporation of deformation effects in DRHBc results in enhanced nuclear binding energies and a notable reduction in $\alpha$-decay energies. With the rotational corrections further incorporated, the theoretical deviation by DRHBc from the experimental data is further reduced. Based on the two-neutron gap $\delta_{\rm 2n}$ and the neutron pairing energy $E_{\rm pair}^n$, prominent shell closures at $N=184$ and $258$, as well as potential sub-shell structures at $N=142, 150, 162, 178, 218$, and $230$ are exhibited. A quasi-periodic variation among prolate, oblate, and spherical shapes as well as prolate deformation predominance have been shown in the evolutions of the quadrupole deformation. Possible shape coexistence is predicted in $^{331}$Bk with the oblate and prolate minima in close energies, which is further supported by the triaxial relativistic Hartree-Bogoliubov theory in continuum~(TRHBc) calculations. The neutron, proton, and charge radii predicted by DRHBc reveal pronounced kink structures at $N=184$ and $258$ in their evolutions with neutron number and compared to those by RCHB, deformation effect significantly enhances the radii of open-shell nuclei.
\end{abstract}

\keywords{odd-$A$ transuranium Bk isotopes, drip line, shell structure, shape evolution and shape coexistence, deformation effects, DRHBc theory}

\maketitle

\section{Introduction}
\label{sec.I}

Element $Z=92$, uranium (U), is the heaviest element that naturally exists on Earth~\cite{Thoennessen2016}. So far, $26$ transuranium elements have been synthesized artificially, ranging from $Z=93$ to $Z=118$~\cite{Thoennessen2016, JPSJ2004Morita_73_2593, Cai2006}, and the $7$th period of the periodic table of elements has been completed. Meanwhile, many theoretical studies predicted the existence of the ``island of superheavy nuclei" near $Z=114$ and $N=184$~\cite{NP1966Myers_81_1, PL1966Wong_21_688, PL1966Sobiczewski_22_500, NPA1969Sobiczewski_131_67, ZAP1969Mosel_222_261}. Thereafter, exploring the island of superheavy nuclei has become as one of the foremost scientific objectives in both experimental research at large-scale facilities and theoretical investigations~\cite{NPR2017Zhou_34_318, JBNU2022Zhang_58_392, CSB2020Yang_65_8, Science2005Seife_309_78}. Extensive investigations on the nuclear structure, decay modes, and reactions have been performed for superheavy nuclei~\cite{PHYSICS2014Zhou_43_817, NPR2014Li_31_253, Lv2016Superheavy,PRC2024He_110_014301, PRC2024Ismail_109_014606, PRC2024Denisov_109_044618}. Among these aspects, the shell structure and magic numbers, the position of the island of superheavy nuclei, as well as the ground state properties such as mass and shape are some key issues for structures of superheavy nuclei which have been achieving great attentions~\cite{NPR2017Zhou_34_318}.

A variety of theoretical models have been developed or employed to investigate the structures and properties of transuranium isotopes and superheavy nuclei~\cite{NPR2014Li_31_253, NPR2013Zhang_30_268}, including several empirical formulas such as the \emph{ab} formula~\cite{SS2012Wen_42_22}, macroscopic-microscopic models~\cite{NPA1994Cwiok_573_356, PRC2011Adamian_84_024324}, configuration-constrained total-Routhian-surface method~\cite{PRL2004Xu_92_252501, PRC2012Liu_86_011301}, self-consistent mean-field models~\cite{PRC2011Adamian_84_024324, PRC1998Bender_58_2126, NPA2006Delaroche_771_103}, quasiparticle phonon model~\cite{JPG2011Jolos_38_115103}, particle-rotor model~\cite{CTP2012Zhuang_57_271}, projected shell model~\cite{PRC2008Sun_77_044307, PRC2008Chen_77_061305, PRC2009Al-Khudair_79_034320}, multishell shell model for heavy nuclei~\cite{PRC2014Cui_90_014321}, cranking shell model~\cite{JPG1982Egido_8_L43, NPA1984Egido_423_93, NPA2009He_817_45}, etc.

The quantum shell effect is the fundamental reason for the existence of superheavy nuclei. Early nuclear structure theories predicted that the atomic nucleus with $Z=114$ and $N= 184$ was doubly magic after $^{208}$Pb. Recent macroscopic-microscopic models still give the same prediction~\cite{PRC2014Mo_90_024320}. However, microscopic models give different predictions. For example, in Ref.~\cite{PRC1997Rutz_56_238}, taking the Skyrme-Hartree-Fock model, Rutz \emph{et al.} predicted magic numbers $Z= 120$ and $N= 172$, $Z= 126$ and $N= 184$ besides $Z= 114$ and $N=184$. In Ref.~\cite{NPA2005Zhang_753_106}, taking the spherical relativistic continuum Hartree-Bogoliubov (RCHB) theory, Zhang \emph{et al.} predicted the proton magic numbers $Z= 120$, $132$, $138$ and neutron magic numbers $N= 172$, $184$, $198$, $228$, $238$, $258$ by examining various physical quantities, such as two-neutron/proton separation energies, two-neutron/proton gaps, shell correction energy, pairing energy, and pairing gap. By further including the deformation, tensor force, and Fock term, different shell structures have been predicted in superheavy nuclei~\cite{JPG2012Zhou_39_085104, PLB2014li_732_169}. For example, in Ref.~\cite{PLB2014li_732_169}, taking the spherical relativistic continuum Hartree-Fock-Bogoliubov model, Li \emph{et al.} predicted proton magic numbers $Z= 120$, $138$ and neutron magic numbers $N= 172$, $184$, $228$, $258$. In Ref.~\cite{NPA1991Patyk_533_132}, taking macroscopic-microscopic models, Patyk and Sobiczewski studied the effects of high-order deformation on nuclear shell structures and revealed $^{270}$Hs being a deformed double magic nucleus. They also pointed out that relatively stable deformed superheavy nuclei exist in the region of $Z= 108$ and $N= 162$, which has been further confirmed by Pei \emph{et al.} through the studies taking the Skyrme-Hartree-Fock$+$BCS model~\cite{NPR2003Pei_20_116}. Besides the important role of nuclear deformation on the shell structures, many studies have proved its significance, such as for single-particle structures~\cite{BOOK1998Bohr_NuclearStructure,PRC2024SunTT}, deformed halos~\cite{PRC2010Zhou_82_011301, JP2011Zhou_312_092067, AIP2012Zhou_1491_208,PRC2020Sun_101_014321,PLB2024SSJLAn}, fission barriers~\cite{CPC2024Zhang_48_104105, PRC2009Moller_79_064304, PS2016Zhou_91_063008}, shape coexistence~\cite{RevModPhys2011, SC2021Chen_64_282011,CPC2022Sun_46_074106}, $\alpha$ decay~\cite{JPG2021Cheng_48_095106} and electron capture cross section~\cite{Nature2020Tsunoda_587_66}.

The deformed relativistic Hartree-Bogoliubov theory in continuum (DRHBc) theory~\cite{PRC2010Zhou_82_011301,PRC2012Li_85_024312,PRC2020Zhang_102_024314}, which can concurrently take into account pairing correlations, continuum effects, and degrees of freedom in deformation, is one of the most potent models. Based on the DRHBc theory, numerous interesting studies have been conducted, such as the prediction of shape-decoupling between the core and halo in deformed halo nuclei~\cite{PRC2010Zhou_82_011301, NPA2020Sun_1003_122011, PRC2021Sun_103_054315, PLB2023Zhang_844_138112,PLB2024Pan}, the resolution of the puzzles regarding the radius and neutron configuration in $^{22}$C~\cite{PLB2018Sun_785_530}, the number of particles in the classically forbidden regions for magnesium isotopes~\cite{PRC2019Zhang_100_034312}, the deformation effects on the neutron drip line~\cite{IJMPE2021In_30_2150009}, the shape evolution, shape coexistence and prolate-shape dominance~\cite{PRC2022Kim_105_034340, PRC2023Guo_108_014319, PRC2023Zhang_108_024310, PLB2023Mun_847_138298}, the fission barrier~\cite{CPC2024Zhang_48_104105}, the evolution of shell closures~\cite{PRC2023Zhang_107_L041303, CPC2024Zheng_48_014107}, the stability peninsulas beyond the neutron drip line~\cite{PRC2021Zhang_104_l021301, PRC2021Pan_104_024331, CPC2021He_45_101001,PRC2024He_110_014301}, the one-proton emission from $^{148-151}$Lu~\cite{PLB2023Xiao_845_138160,PLB2024SSZhang}, the dependence on the multipole expansion order~\cite{IJMPE2019Pan_28_1950082}, the optimization of Dirac-Woods-Saxon~(DWS) basis~\cite{PRC2022Zhang_106_024302}, the rotational excitations of exotic nuclei with angular momentum projection~\cite{PRC2021Sun_103_054315, SB2021Sun_66_2072}, and the dynamical correlation with a two-dimensional collective Hamiltonian method~\cite{CPC2022Sun_46_064103}. 

With a proton number of $Z = 97$, Bk is located almost in the middle between the proton shell of $Z = 82$ and the possible proton shells $Z = 114, 120, 126$. From an experimental perspective, it is proposed that $^{249}$Bk is an ideal target for synthesizing elements $119$ and $120$~\cite{PRC2020Khuyagbaatar_102_064602}. However, a systematic and detailed research on the properties of Bk isotopes is still absent and urgently desired. In this paper, we will systematically investigate the ground-state properties and structure evolutions of odd-$A$ transuranium Bk isotopes. Matters such as the determination of drip lines, shell structures and magic numbers, the impact of deformation on binding energies and radii, shape evolution and shape coexistence will be inspected. This paper is organized as follows: the DRHBc theory will be succinctly introduced in Section~\ref{sec.II}, after presenting the results and discussions in Section~\ref{sec.III}, finally a summary will be offered in Section~\ref{sec.IV}.

\section{Theoretical framework and numerical details}
\label{sec.II}

Detailed descriptions of the DRHBc theory can be found in Refs.~\cite{PRC2010Zhou_82_011301, CPL2012Li_29_042101, PRC2020Zhang_102_024314}. Here, we briefly introduce the formalism for the convenience of discussions. In the DRHBc theory, the relativistic Hartree-Bogoliubov (RHB) equation reads,

\begin{equation}
   ~\begin{pmatrix}
     ~\hat{h}_D-\lambda_\tau &&~\hat{\Delta}\\
      -\hat{\Delta}^* && -\hat{h}_D^*+\lambda_\tau
   ~\end{pmatrix}
   ~\begin{pmatrix}
      U_k\\
      V_k
   ~\end{pmatrix}
    =E_k
   ~\begin{pmatrix}
      U_k\\
      V_k
   ~\end{pmatrix},
\end{equation}
where $\hat{h}_D$ represents the Dirac Hamiltonian, $\hat{\Delta}$ is the pairing potential, $\lambda_\mathrm{\tau}$ is the Fermi energy for neutrons or protons $(\tau=n,p),\:E_{k}$ is the quasiparticle energy, and $U_k$ and $V_k$ are the quasiparticle wave functions.

The Dirac Hamiltonian in the coordinate space is given by
\begin{equation}
  h_{D}({\boldsymbol{r}})={\bm~\alpha}\cdot{\boldsymbol{p}}+V({\boldsymbol{r}})+\beta[M+S({\boldsymbol{r}})],
\end{equation}
where $M$ is the nucleon mass, and $S({\bm r})$ and $V({\bm r})$ are the scalar and vector potentials, respectively. The pairing potential is expressed as,
\begin{equation}
 ~\Delta(\boldsymbol{r}_{1},~\boldsymbol{r}_{2})=V^{pp}\left(\boldsymbol{r}_{1},~\boldsymbol{ r}_{2}\right)\kappa(\boldsymbol{r}_{1},~\boldsymbol{r}_{2}),
\end{equation}
where $\kappa=V^*U^T$ is the pairing tensor and $V^{pp}$ is the pairing force of a density-dependent zero-range type,
\begin{equation}
  V^{pp}\left(\boldsymbol{r}_{1},\boldsymbol{r}_{2}\right)=V_{0}\frac{1}{2}\left(1-P^{\sigma}\right)\delta\left(\boldsymbol{r}_{1}-\boldsymbol{r}_{2}\right)\left(1-\frac{\rho\left(\boldsymbol{r}_{1}\right)}{\rho_{\mathrm{sat}}}\right),
\end{equation}
with $V_0$ being the pairing strength, $\frac{1}{2}(1-P^\sigma)$ the projector for the spin $S=0$ component, and $\rho_{\rm sat}$ the saturation density of nuclear matter.

For an axially deformed nucleus with spatial reflection symmetry, the third component $K$ of the angular momentum $j$ and the parity $\pi$ are conserved quantum numbers. Thus, the RHB Hamiltonian can be decomposed into blocks $K^\pi$ characterized by $K$ and parity $\pi$.

Additionally, the potentials and densities in the DRHBc theory can be expanded in terms of Legendre polynomials~\cite{PRC1987Price_36_354},
\begin{equation}
  f(\boldsymbol{r})=\sum_{\lambda}f_{\lambda}(r)P_{\lambda}(\cos\theta),\quad\lambda=0,2,4,\cdots,
 ~\label{Eq:Legend}
\end{equation}
with
\begin{equation}
  f_\lambda(r)=\frac{2\lambda+1}{4\pi}\int d\Omega f(\boldsymbol{r})P_\lambda(\cos\theta).
\end{equation}

For an odd-$A$ or odd-odd nucleus, one needs to further take into consideration the blocking effect for the unpaired single proton or neutron~\cite{PLB2018Sun_785_530, IJMPE2021In_30_2150009, PRC2008Perez-Martin_78_014304}. The equal filling approximation is adopted to deal with the blocking effects in the DRHBc theory~\cite{PRC2022Pan_106_014316}.

The RHB equations are solved in the DWS basis~\cite{PRC2003Zhou_68_034323, PRC2012Li_85_024312, PLB2023Mun_847_138298, CPC2022Wang_46_024107}, which can appropriately describe the large spatial extension of weakly bound nuclei. In the numerical calculation, the angular momentum cutoff for the DWS basis is chosen as $J_{\max}=\frac{23}{2}h$. The maximum expansion order in Eq.~(\ref{Eq:Legend}) is $\lambda_{\max}=8$, which is sufficient for our study~\cite{PRC2022Pan_106_014316, PRL2009Nortershauser_102_062503}. The size of the box is set to be $20$~fm, and the energy cutoff in the Fermi sea is $E_\mathrm{cut}^+=300$ MeV. For the particle-particle channel, we use the zero-range pairing force with a saturation density $\rho_{\mathrm{sat}}= 0.152\mathrm{~fm}^{- 3}$ and a pairing strength $V_0=-325$~MeV$\cdot$fm$^{3}$~\cite{ADNDT2022Zhang_144_101488,IJMPE2021In_30_2150009}. All the numerical details are the same as those employed in constructing the DRHBc mass tables~\cite{ADNDT2022Zhang_144_101488,ADNDT2024Guo}.

\section{Results and discussion}
\label{sec.III}

In this section, to reveal the ground-state properties and the structure evolutions of the odd-$A$ transuranium Bk isotopes, systematic calculations have been carried out from the proton drip line to the neutron drip line using the DRHBc theory. The results are further compared with the available experimental data~\cite{CPC2021Wang_45_030003} and those from spherical RCHB calculations~\cite{ADNDT2018Xia_121_1}. In Part A, quantities related to nuclear masses such as binding energy, two-neutron and two-proton separation energies, decay energy, Fermi energy, two-neutron gap, and pairing energy are given, based on which the shell structures have been discussed. In Part B, the evolution of nuclear shapes and possible shape coexistence are examined in the Bk isotopic chain. Finally, in Part C, the density distributions are shown, and the root-mean-square (rms) neutron, proton, and charge radii are predicted.

\subsection{Nuclear masses and shell structures \label{subsec.I}}

\begin{figure}[t!]
\centering
\includegraphics[width=0.85\linewidth]{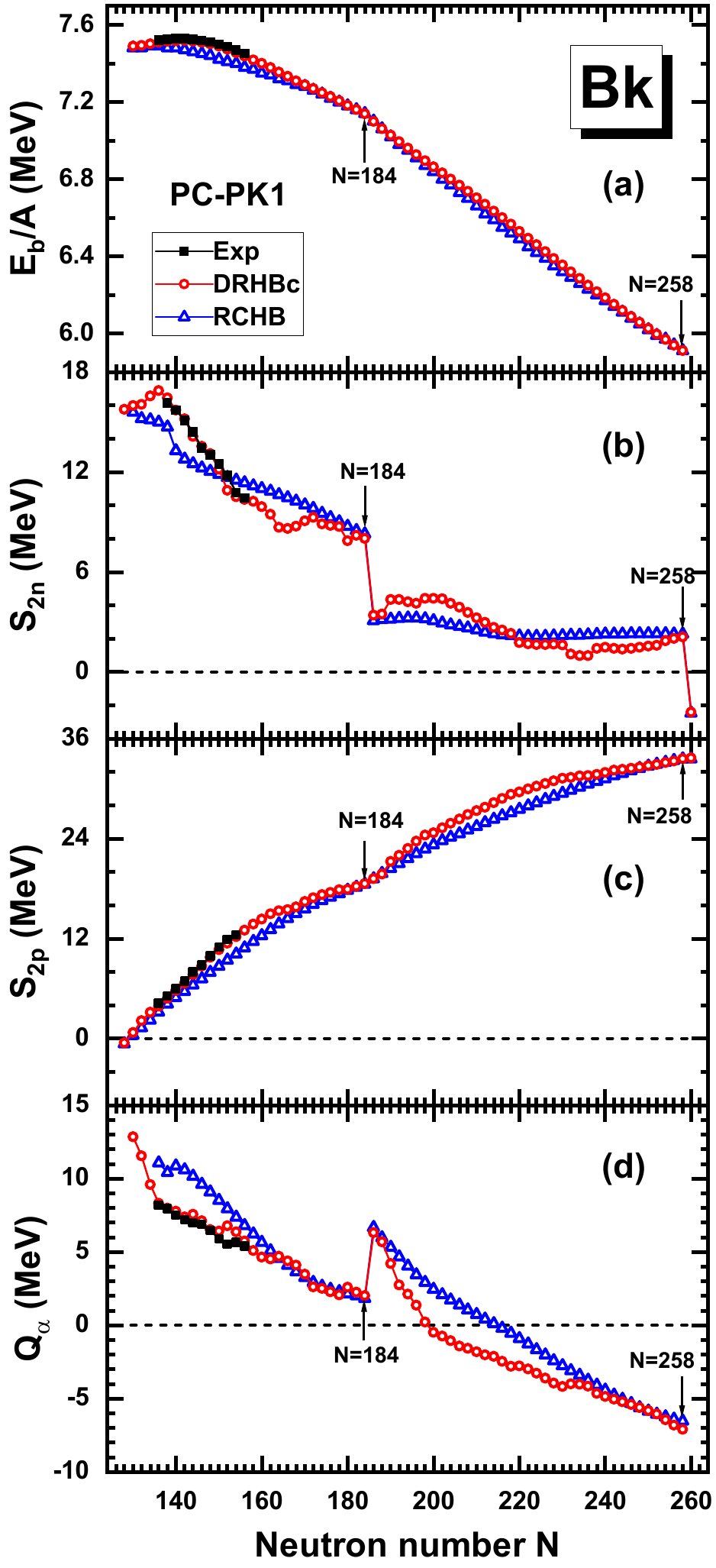}
\caption{(Color online) (a) Binding energy per nucleon $E_{\rm b}/A$, (b) two-neutron separation energy $S_{\rm 2n}$, (c) two-proton separation energy $S_{\rm 2p}$, and (d) $\alpha$ decay energy $Q_{\alpha}$ as functions of the neutron number for the odd-$A$ Bk isotopes obtained in the DRHBc calculations with PC-PK1. The results by the RCHB calculations~\cite{ADNDT2018Xia_121_1} and the available experimental data~\cite{CPC2021Wang_45_030003} are presented for comparison.}
\label{Fig1}
\end{figure}

In Fig.~\ref{Fig1}(a), the binding energies per nucleon $E_{\rm b}/A$ for the odd-$A$ Bk isotopes are depicted as a function of the neutron number $N$ and compared with the RCHB results and the available data. Within the range of available experimental data, the calculation outcomes by DRHBc are in good accordance with the experimental data, whereas the results by RCHB deviate significantly from the experimental results. This implies that deformation effects have a considerable influence on the binding energies of those nuclei. The nucleus $^{239}$Bk has the largest specific binding energy experimentally, which is reproduced by the DRHBc calculations. When comparing with the RCHB results, the DRHBc gives very similar specific binding energies near the possible neutron shell closures, namely $N=184$ and $N=258$, while visible differences are exhibited when deviating the magic neutron numbers, especially at the mid-shell regions.
 
Fig.~\ref{Fig1}(b) presents the two-neutron separation energies $S_{\rm 2n}(Z,N)=E_{\rm b}(Z,N)-E_{\rm b}(Z,N-2)$ as a function of the neutron number $N$ by both DRHBc and RCHB theories, in comparison with the experimental data~\cite{CPC2021Wang_45_030003}. For both DRHBc and RCHB calculations, $S_{\rm 2n}$ gradually decreases with the increasing neutron number, drops sharply at $N=184$, then remains about $3$~MeV within a large mass range until it drops sharply again at $N=258$, and finally becomes negative at $N=260$, indicating that $^{355}$Bk is the neutron drip-line nucleus for the Bk isotopes. This evolution also demonstrates that $N=184$ and $258$ are two magic numbers for neutrons. Compared to the RCHB calculations, DRHBc reproduces the available experimental data much better. Furthermore, remarkable crossings of $S_{\rm 2n}$ within two shell closures can be found between the results by DRHBc and RCHB. For instance, from $N=184$ to $258$, the $S_{\rm 2n}$ values by DRHBc are larger than those by RCHB for nuclei before mid-shell, and the opposite holds true after that. The main reason can be attributed to the deformation effect encompassed in DRHBc calculations. The difference of $S_{\rm 2n}$ between the DRHBc and RCHB can be delineated as follows,
\begin{eqnarray}
\Delta S_{\rm 2n}(Z,N)&=&S_{\rm 2n}({\rm DRHBc})-S_{\rm 2n}({\rm RCHB})\nonumber\\
&=&\Delta E_{\rm b}(Z,N)-\Delta E_{\rm b}(Z,N-2)\nonumber\\
&=&2\frac{\partial \Delta E_{\rm b}}{\partial N}:\propto \frac{\partial \Delta|\beta_2|}{\partial N}\equiv\frac{\partial |\beta_2|}{\partial N},%
\label{Eq:DeltaS2n}%
\end{eqnarray}%
where $\Delta E_{\rm b}$ and $\Delta|\beta_{2}|$ denote the difference between the DRHBc and RCHB results for $E_{\rm b}$ and $|\beta_{2}|$, respectively. Due to the spherical assumption in RCHB, $\Delta|\beta_{2}|$ is always equal to $|\beta_2|$ obtained by DRHBc. The symbol $:\varpropto$ represents that the $\Delta E_{\rm b}$ strongly correlates with $\Delta|\beta_{2}|$, i.e., the gain in binding energy due to deformation, expressed by the energy difference $\Delta E_{\rm b}$, is consistent with the absolute value of the deformation parameter as discussed in Ref.~\cite{IJMPE2021In_30_2150009}. In general, as the nucleon number moves away from the shell closure to the mid-shell, deformed structures are developed while suppressed when nucleon number increases from the mid-shell to the next shell closure. According to this kind of deformation evolution within two shell closures, $\frac{\partial |\beta_2|}{\partial N}>0$ before mid-shell, $\frac{\partial |\beta_2|}{\partial N}<0$ after mid-shell and $\frac{\partial |\beta_2|}{\partial N}\approx 0$ around mid-shell. Consequently, $S_{\rm 2n}$ in DRHBc are greater than those in RCHB for nuclei before mid-shell while smaller after mid-shell. Around the mid-shell, the $S_{\rm 2n}$ values in DRHBc and RCHB are close to each other which then results in the crossings of $S_{\rm 2n}$ in Fig.~\ref{Fig1}(b). In addition, as discussed in Ref.~\cite{IJMPE2021In_30_2150009}, the inclusion of deformation may extend or shrink the position of drip line depending on the evolution of the degree of deformation. For Bk isotopic chain, due to spherical shapes for nuclei around the drip line, the identical positions of the drip line are predicted by DRHBc and RCHB.

Fig.~\ref{Fig1}(c) presents the two-proton separation energies $S_{\rm 2p}(Z,N)=E_{\rm b}(Z,N)-E_{\rm b}(Z-2,N)$ as a function of the neutron number $N$. The $S_{\rm 2p}$ increases with the increasing neutron number and presents kink structures at the shell closures of $N=184$ and $258$. Similar to $S_{\rm 2n}$, the difference of $S_{\rm 2p}$ between the DRHBc and RCHB calculations reads, 
\begin{eqnarray}
\Delta S_{\rm 2p}(Z,N)&=&S_{\rm 2p}({\rm DRHBc})-S_{\rm 2p}({\rm RCHB})\nonumber\\
&=&\Delta E_{\rm b}(Z,N)-\Delta E_{\rm b}(Z-2,N)\nonumber\\
&\approx&2\frac{\partial \Delta E_{\rm b}}{\partial Z}:\propto \frac{\partial \Delta|\beta_2|}{\partial Z}\equiv\frac{\partial |\beta_2|}{\partial Z},
\label{Eq:DeltaS2p}
\end{eqnarray}
which means the sign of $\Delta S_{\rm 2p}(Z,N)$ is determined by $\frac{\partial |\beta_2|}{\partial Z}$. Since the Bk~($Z=97$) isotopes are more deformed than the corresponding Am~($Z=95$) isotones as seen in Ref.~\cite{Particle2026WZhang}, the $\Delta S_{\rm 2p}(Z,N)$ will be positive. As a result, compared with the RCHB results, the DRHBc predicts larger $S_{\rm 2p}$ for Bk isotopes as shown in Fig.~\ref{Fig1}(c). In addition, on the neutron-deficient side, $S_{\rm 2p}$ changes from positive to negative from $N=130$ to $128$ by both the DRHBc and RCHB theories, indicating the same proton drip-line nucleus $^{227}$Bk. 

\begin{figure}[t!]
\centering
\includegraphics[width=0.85\linewidth]{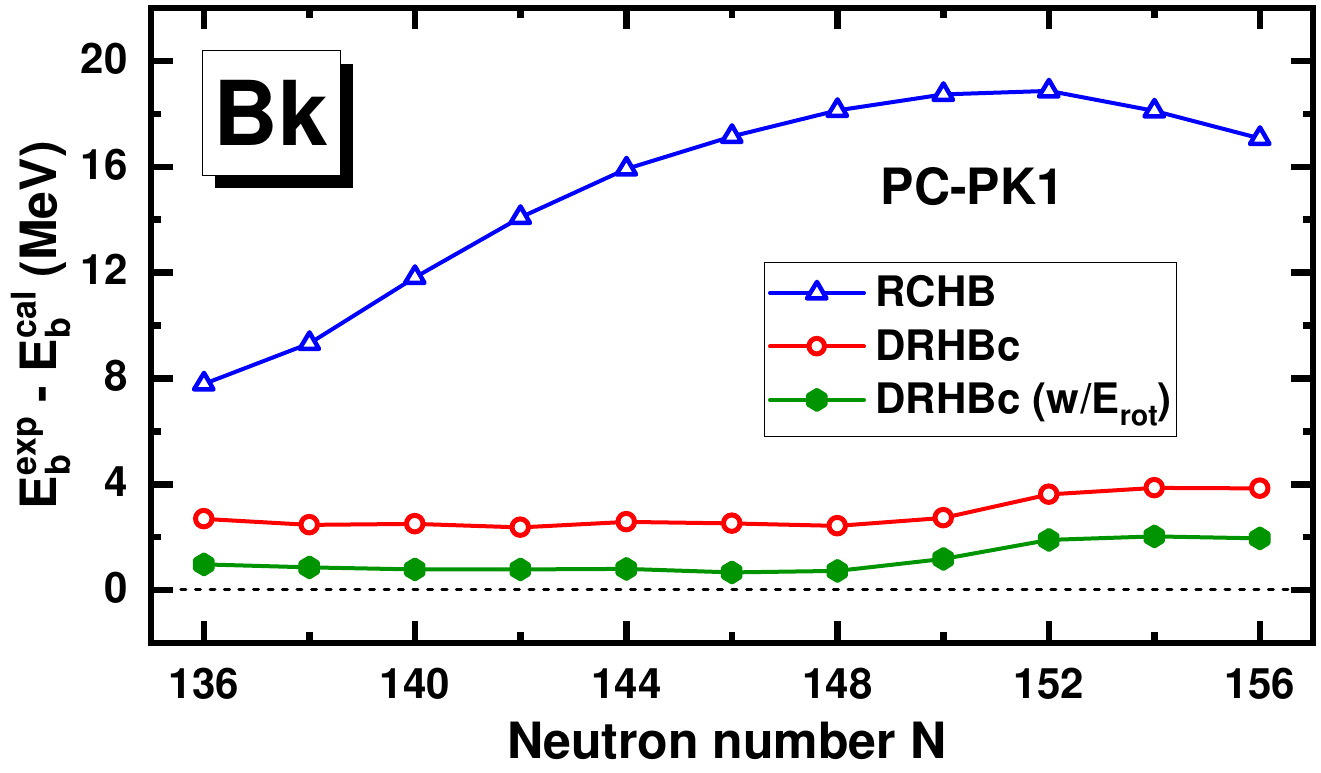}
\caption{(Color online) The difference of the total binding energy $E_{\rm b}^{\rm exp}$-$E_{\rm b}^{\rm cal}$ between the experimental data and the DRHBc results with and without the rotation correction $E_{\mathrm{rot}}$. The RCHB results~\cite{ADNDT2018Xia_121_1} are also provided for comparison.}
\label{Fig2}
\end{figure}

Fig.~\ref{Fig1}(d) presents the $\alpha$-decay energies $Q_\alpha(Z,N)=E_{\rm b}(Z-2,N-2)+E_{\rm b}(2,2)-E_{\rm b}(Z,N)$ as a function of the neutron number $N$. With the inclusion of deformation, the DRHBc results are in good consistency with the experimental data. In general, the $Q_{\alpha}$ decreases gradually with the neutron number $N$ with a sudden increase at the shell closure of $N=184$, and then becomes negative at $N=200$. Compared with the RCHB results, $Q_\alpha$ values in DRHBc predications get significantly reduced for Bk isotopes before mid-shell. This behavior can be understood by the $S_{\rm 2n}$ and $S_{\rm 2p}$ shown in Figs.~\ref{Fig1}(b) and \ref{Fig1}(c). In fact, $Q_{\alpha}$ can be written in a form of $S_{\rm 2n}$ and $S_{\rm 2p}$ as $Q_\alpha(Z,N)=E_{\rm b}(2,2)-S_{\rm 2p}(Z,N-2)-S_{\rm 2n}(Z,N)$. The difference of $Q_\alpha$ between DRHBc and RCHB results is the consequence of the crossings of $S_{\rm 2n}$ in Fig.~\ref{Fig1}(b) and the larger $S_{\rm 2p}$ of the DRHBc in Fig.~\ref{Fig1}(c). It is noted that the last Bk isotope predicted to have $\alpha$ decay by the DRHBc theory ($N=198$) has $16$ fewer neutrons than that by RCHB ($N=214$).
 
In Fig.~\ref{Fig2}, the difference of the total binding energy between the experimental data and the DRHBc results $E_{\rm b}^{\rm exp}$-$E_{\rm b}^{\rm cal}$ are plotted, in comparison with the RCHB results~\cite{ADNDT2018Xia_121_1}. We further take the rotational correction $E_{\rm rot}$ (for the details, see Ref.~\cite{PRC2012Li_85_024312,PRC2020Zhang_102_024314}) into account for the total binding energies $E_{\rm b}$ in the DRHBc calculations. As already discussed in Fig.~\ref{Fig1}(a), compared with the RCHB results~\cite{ADNDT2018Xia_121_1}, the DRHBc calculations have already prominently improved the predicted binding energies with the deformation effects being considered. With the rotational corrections further incorporated, the theoretical deviation from the experimental data is further reduced by almost $2$~MeV to a root-mean-square error of $1.23$~MeV. Hence, it can be concluded that the rotational corrections also have considerable impacts on the binding energies of Bk isotopes.

\begin{figure}[t!]
\centering
\includegraphics[width=0.85\linewidth]{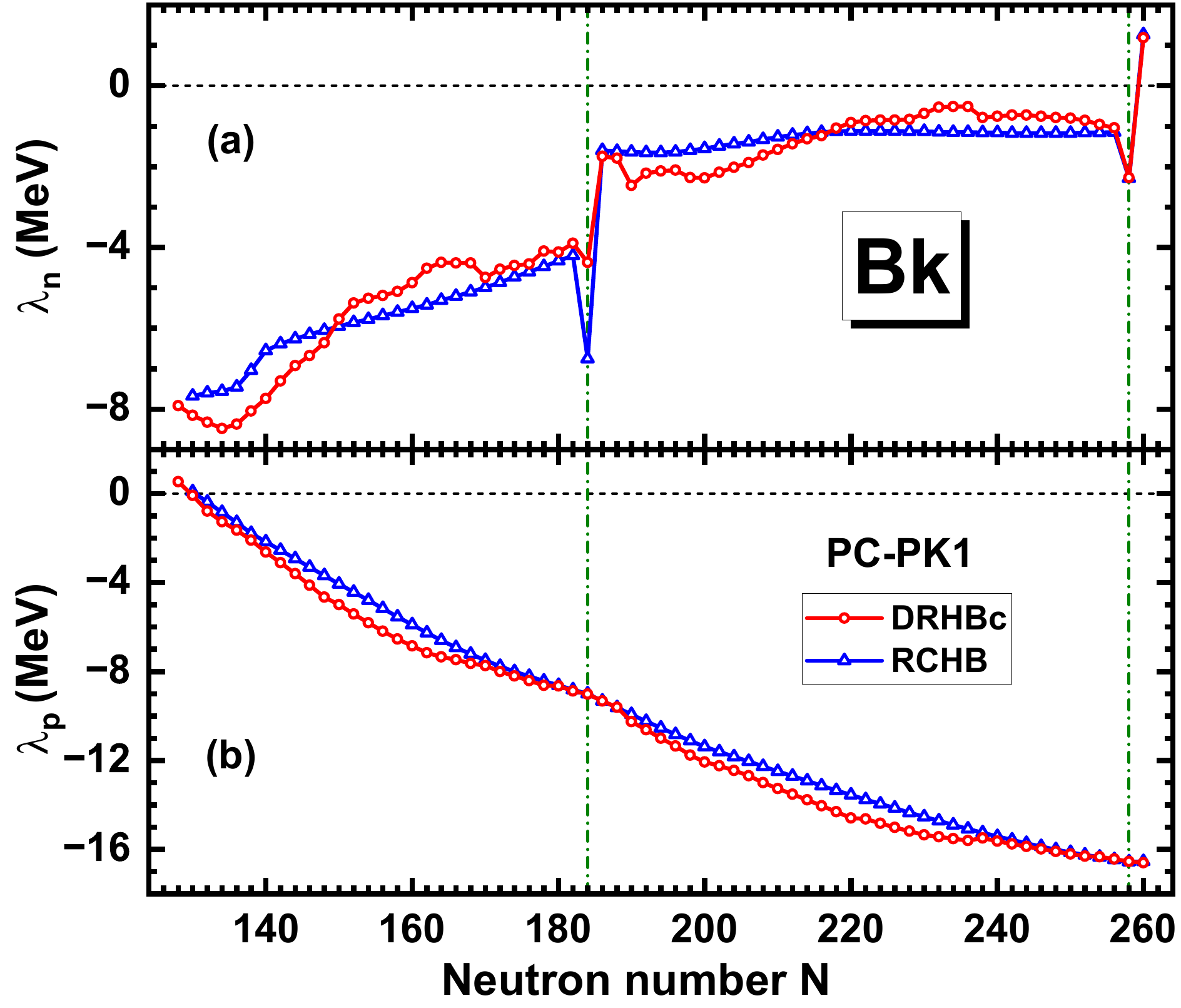}
\caption{(Color online) The (a) neutron and (b) proton Fermi energies of the Bk isotopes obtained by the DRHBc calculations as functions of the neutron number. The RCHB results in Ref.~\cite{ADNDT2018Xia_121_1} are also presented for comparison.}
\label{Fig3}
\end{figure}

Within the mean-field framework, the Fermi energy represents the variation of the total energy with respect to the particle number and conveys information regarding the drip line position. Fig.~\ref{Fig3}(a) presents the neutron Fermi energy $\lambda_{\rm n}$ as a function of the neutron number $N$ by both the DRHBc and RCHB theories. In both DRHBc and RCHB calculations, $\lambda_{\rm n}$ rises continuously with neutron number and a remarkable increase happens at shell closure of $N=184$. It surpasses the continuum threshold of $0~$MeV at $N=260$, indicating the neutron drip-line nucleus being $^{355}$Bk, which is consistent with the prediction by the criterion of $S_{\rm 2n}$ shown in Fig.~\ref{Fig1}(b). It is noted that $\lambda_{\rm n}$ for $^{281}$Bk and $^{355}$Bk are given by the last occupied single-proton level due to the pairing collapse. Crossings of $\lambda_{\rm n}$ within two shell closures can be found between the results by DRHBc and RCHB. For instance, from $N=184$ to $258$, the $\lambda_{\rm n}$ values by DRHBc are smaller than those by RCHB for nuclei before mid-shell while the opposite holds true after the mid-shell. This can be accounted for by the different single-particle~(s.p.)~level structures in the spherical and deformed cases. The s.p.~levels with the same total angular momentum $j$ are degenerate when spherical and will split into $(2j+1)/2$ levels with the third component $K=1/2, 3/2,\cdots, j/2$ when quadruple deformed. The last occupied level in the deformed case is lower than that in the spherical case for nuclei before mid-shell while opposite after mid-shell.
 
Fig.~\ref{Fig3}(b) presents the proton Fermi energy $\lambda_{\rm p}$ for Bk isotopes as a function of neutron number $N$. With the increasing neutron number $N$, as the mean-field potential for protons will be deepen, the value of $\lambda_{\rm p}$ decreases correspondingly by both RCHB and DRHBc. Besides, kink structures are presented at the shell closures of $N=184$ and $258$. Compared with RCHB results, $\lambda_{\rm p}$ predicted by DRHBc are smaller generally. This can be explained by the different proton s.p. structures in the spherical and deformed cases, similar to the case of $\lambda_{\rm n}$ in Fig.~\ref{Fig3}(a). As the Bk isotopes with $Z=97$ are positioned before the mid-shell between proton shell closures of $Z=82$ and $126$, the last occupied proton level in the deformed case is lower than that in the spherical case. In addition, on the neutron-deficient side, $\lambda_{\rm p}$ surpasses the continuum threshold of $0~$MeV at $N=128$, indicating $^{227}$Bk as the proton drip-line nucleus in DRHBc, which is the same as that predicted according to $S_{\rm 2p}$ in Fig.~\ref{Fig1}(c). The proton drip-line nucleus in the RCHB prediction is $^{229}$Bk with two more neutrons.

\begin{figure}[t!]
\centering
\includegraphics[width=0.85\linewidth]{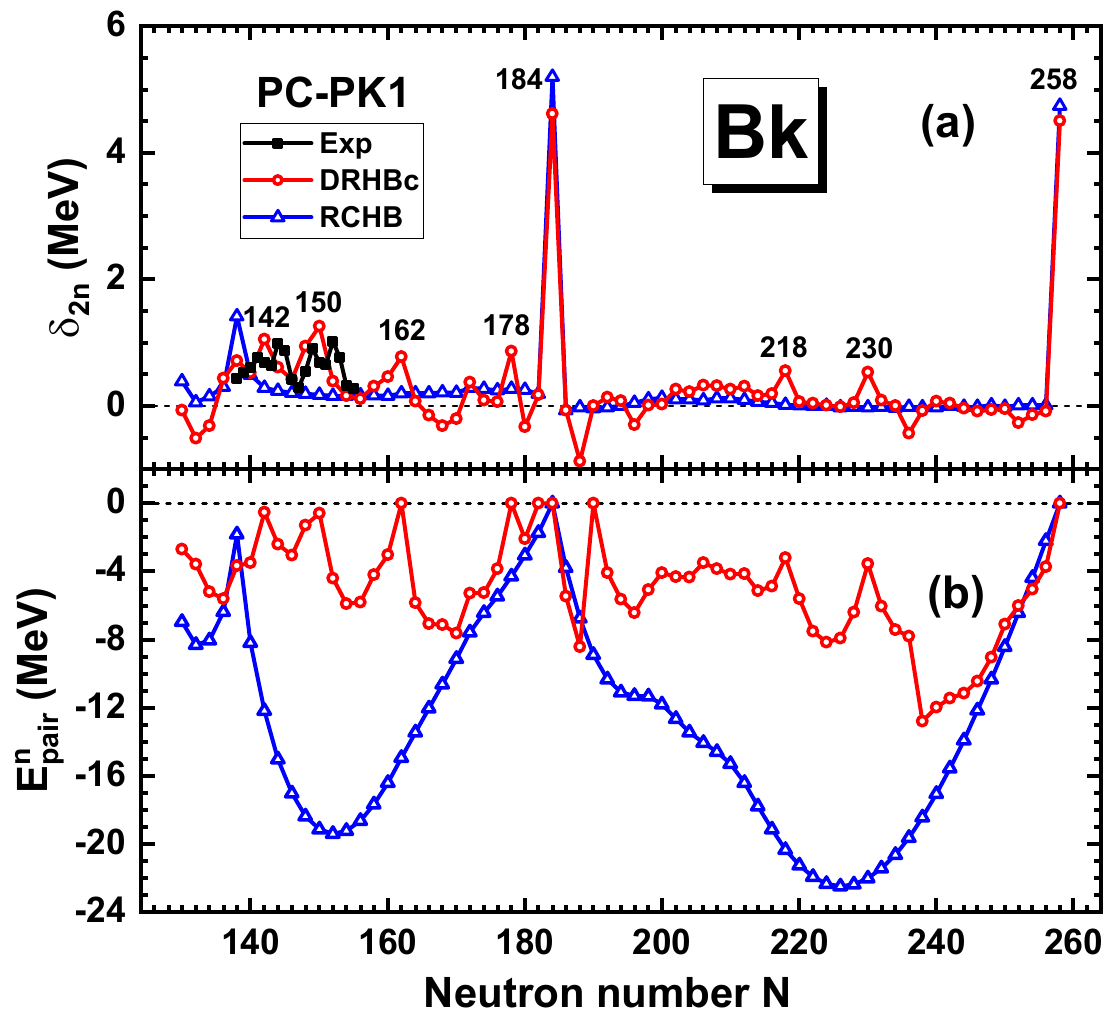}
\caption{(Color online) (a) Two-neutron gap $\delta_{\rm 2n}$ and (b) neutron pairing energy $E^{\rm n}_{\rm pair}$ as functions of the neutron number in the Bk isotopes obtained by the DRHBc calculations. The RCHB results~\cite{ADNDT2018Xia_121_1} and the available experimental data~\cite{CPC2021Wang_45_030003} are also presented for comparison.}
\label{Fig4}
\end{figure}

Compared with the $S_{\rm 2n}$ in Fig~\ref{Fig1}(b), the two-neutron gap $\delta_{\rm 2n}(Z,N)=S_{\rm 2n}(Z,N)-S_{\rm 2n}(Z,N+2)$ is more sensitive to the shell effects and thus it is often employed to search for the shell closures~\cite{NPA2005Zhang_753_106}. Fig~\ref{Fig4}(a) presents $\delta_{\rm 2n}$ for Bk isotopes as a function of the neutron number $N$ by DRHBc, in comparison with the RCHB results and the available data. In general, the DRHBc results agree better with the available data~\cite{CPC2021Wang_45_030003} compared with the RCHB results. 
At $N=184$ and $258$, the $\delta_{\rm 2n}$ exhibits very sharp peaks, indicating significant shell closures there. It is noted that, in the DRHBc predictions, many small peaks also emerge, respectively at $N=142, 150, 162, 178, 218$, and $230$, corresponding to possible sub-shells. In contrast, only one additional peak appears at $N=138$ in the RCHB predictions. Since the pairing energy can also reveal nuclear shell structures, in Fig.~\ref{Fig4}(b), we present the neutron pairing energies $E_{\rm pair}^{\rm n}$ as a function of the neutron number $N$. In RCHB calculations, the $E_{\rm pair}^{\rm n}$ shows very obvious large shell structures: the $|E_{\rm pair}^{\rm n}|$ approaches zero at shell closures of $N=184$, and $258$ or local minimum at $N=138$ while maxima at mid-shells. In DRHBc calculations, many more peaks appear with the $|E_{\rm pair}^{\rm n}|$ approaching zero, respectively at $N=142, 150, 162, 178, 182, 184, 190, 258$, or local minima at $N=218, 230$, which are almost in one-to-one correspondence with the peaks of $\delta_{\rm 2n}$. All those indicated by $\delta_{\rm 2n}$ and $E_{\rm pair}^{\rm n}$ clearly signify the shell structures of Bk isotopes. However, the neutron shells predicted for Bk isotopes by DRHBc might differ from those in the superheavy nuclei with proton numbers $Z=117-120$, where $N=172$ is also predicted to be a large neutron shell, while the shell closure $N=184$ vanishes in $Z=119, 120$~\cite{PRC2024JMYao}.

\subsection{Shape evolution and shape coexistence}
\label{subsec.II}

\begin{figure}[ht!]
\centering
\includegraphics[width=0.85\linewidth]{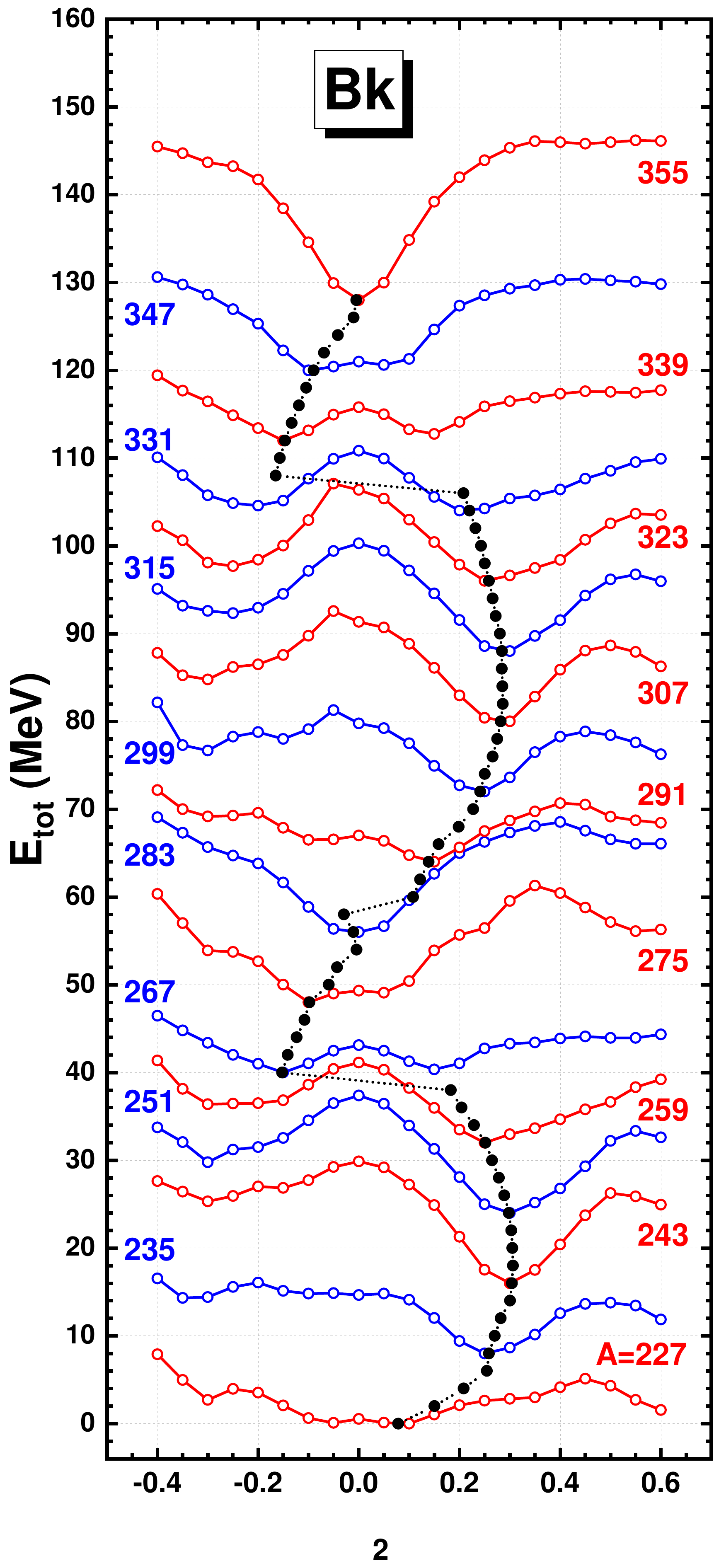}
\caption{(Color online) Evolutions of the potential energy curves (PECs) denoted by open circles of $^{227, 235, \cdots, 355}$Bk isotopes with $\Delta N=8$ obtained by the constrained DRHBc calculations with a step of deformation $\Delta\beta_2 = 0.05$. The PECs have been scaled to the energy of their ground states and shifted upward by 1 MeV for every one neutron. The ground-state deformations (solid circles) by unconstrained DRHBc calculations of odd-$A$ Bk isotopes are also plotted.}
\label{Fig5}
\end{figure}

\begin{figure*}[t!]
\centering
\includegraphics[width=0.95\linewidth]{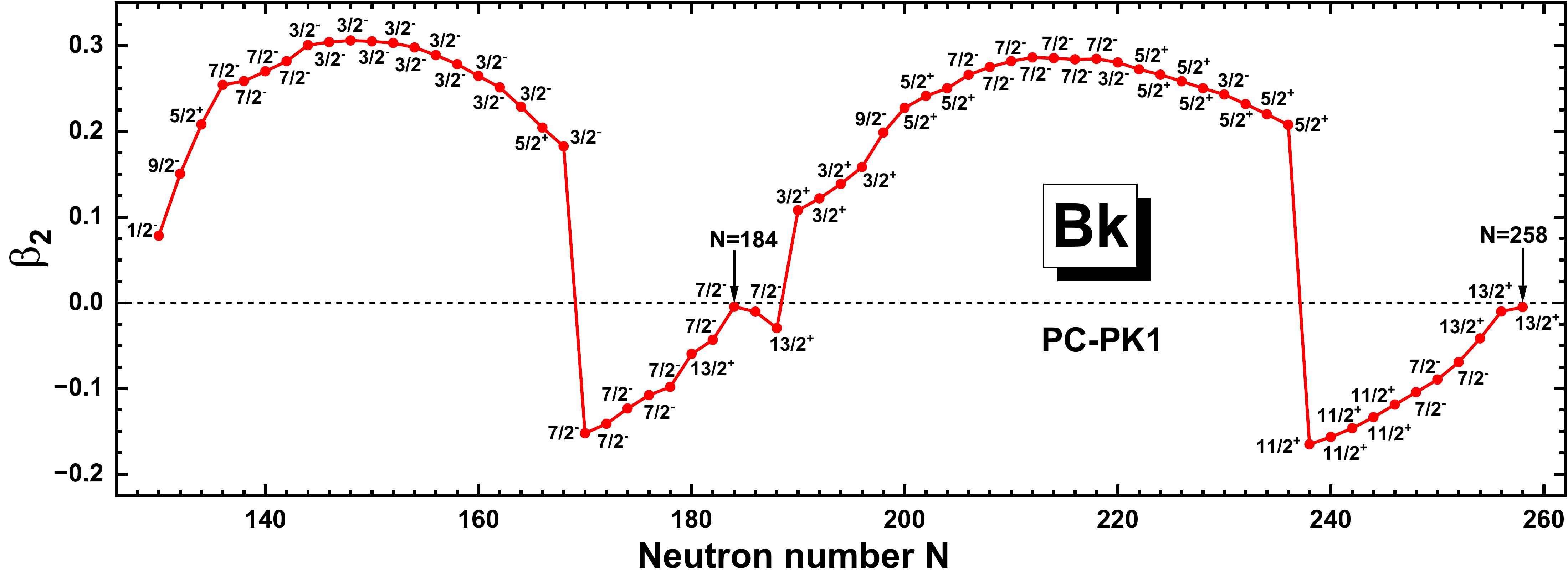}
\caption{ (Color online) Evolution of the quadrupole deformation $\beta_2$ and the spin-parity $K^\pi$ as functions of the neutron number $N$ for the odd-$A$ Bk isotopes in ground states.}
\label{beta2}
\end{figure*}

In Fig.~\ref{Fig5}, the evolutions of the potential energy curves (PECs) by DRHBc are plotted for the Bk isotopes with $\Delta N=8$ ranging from the proton drip line $^{227}$Bk to the neutron drip line $^{355}$Bk. Constrained calculations are done with a step of deformation $\Delta\beta_2 = 0.05$. For comparison, the ground-state deformations by unconstrained calculations are also plotted. For the proton drip-line nucleus $^{227}$Bk, the PEC is very flat and a prolate ground state around $\beta_2 = 0.10$ is determined by both the constrained and unconstrained calculations. In a large mass range from $^{235}$Bk to $^{259}$Bk, distinct prolate global minima are predicted, which are much deeper than the local oblate minima. Instead, in $^{267}$Bk, two minima in prolate and oblate sides have very close energies with the difference as small as $0.35$~MeV and a barrier height of $3.1$~MeV between them, indicating the possible shape coexistence therein. The competition between the two energy minima ultimately determines the ground state in an oblate shape. The nuclear ground state remains oblate for several Bk isotopes until $^{283}$Bk, which has only one energy minimum in spherical due to the neutron shell closure of $N = 184$. After the neutron number exceeds the magic number $N = 184$, the Bk isotopes again become prolate in a large mass range until $^{333}$Bk. This prolate predominance~\cite{BOOK1998Bohr_NuclearStructure} has also been observed in other isotopes~\cite{Casten2000nuclear, PRC2022Sugawara_106_024301, PRC2023Guo_108_014319}. As shown in Fig.~\ref{Fig5}, the PEC of $^{331}$Bk has two energy minima with an energy difference of $0.57$~MeV and a barrier of $6.8$~MeV between them, presenting as another candidate for the possible shape coexistence. The ground state of $^{333}$Bk is in prolate deformation while it turns to be oblate for $^{335}$Bk. The oblate shape remains before the neutron number reaching the neutron shell closure of $N=258$, where $^{355}$Bk is spherical.

To show the shape evolution more intuitively, Fig.~\ref{beta2} presents the ground state deformation $\beta_2$ as a function of the neutron number $N$ obtained by the unconstrained DRHBc calculations. From the proton drip line $^{227}$Bk to the neutron drip line $^{355}$Bk, the nuclear shape undergoes prolate, oblate, and spherical transitions between two shell closures, which is similar as depicted in the uranium isotopes~\cite{CPC2024Zhang_48_104105}. In general, the shape is spherical at $N=184, 258$, while deformed when deviating from the two predicted shell closures. Obvious prolate predominance has been noticed in large mass ranges with $134\leqslant N\leqslant 168$ and $190\leqslant N\leqslant 236$, while nuclear shapes are oblate only in a short mass range before the neutron number reaching the shell closures. Besides, there are some regions with prolate-to-oblate transition, where shape coexistence may occur. According to the shape evolution described by DRHBc, the rate of deformation, represented by $\frac{\partial |\beta_2|}{\partial N}$ in Eq.~(\ref{Eq:DeltaS2n}), evolves from positive to negative as the neutron number exceeds the mid-shell within two shell closures, while remaining zero around the mid-shell. This confirms the crossings of $S_{\rm 2n}$ between the DRHBc and RCHB results as shown in Fig.~\ref{Fig1}(b). 

Figure~\ref{beta2} also illustrates the spin-parity $K^\pi$ of Bk isotopes in their ground states, which is determined by the occupied orbital of the odd proton in the automatic blocking~\cite{PRC2022Pan_106_014316}. As seen, although the proton number of Bk isotopes is fixed to be $97$, the blocked orbital of the odd proton evolves with the change of deformation. In the prolate-dominated region, the spin-parity of Bk isotopes primarily resides in the $3/2^-$, $7/2^-$, $3/2^+$, and $5/2^+$ states, which originate from the spherical $2f_{7/2}$, $1h_{9/2}$, $1i_{13/2}$, and $1i_{13/2}$ orbital, respectively. In contrast, in the oblate region, the predominant spin-parity states are $7/2^-$, $11/2^+$, and $13/2^+$, derived from the spherical $2f_{7/2}$, $1i_{13/2}$, and $1i_{13/2}$ orbital, respectively. Additionally, in the proton-drip line region, $K^\pi$ is observed to be $1/2^-$ for $^{227}$Bk and $9/2^-$ for $^{229}$Bk, while after the neutron shell closure at $N=184$, $K^\pi = 9/2^-$ for $^{295}$Bk. These assignments originate from the spherical $2f_{7/2}$, $1h_{9/2}$, and $1h_{9/2}$ orbital, respectively.

\begin{figure}[t!]
\centering
\includegraphics[width=0.95\linewidth]{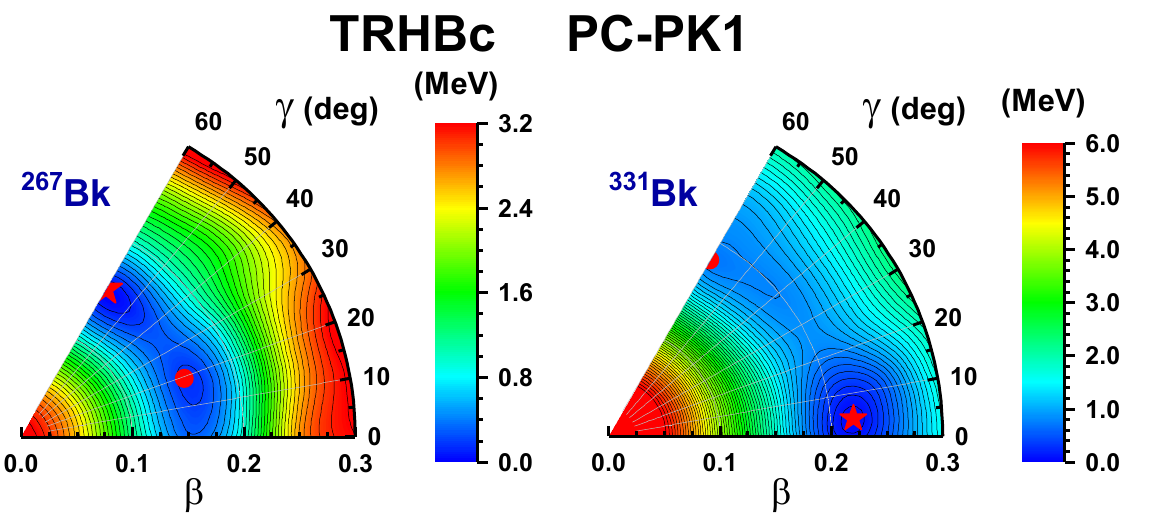}
\caption{(Color online) Potential energy surfaces (PESs) in the $(\beta, \gamma)$ plane for $^{267}$Bk and $^{331}$Bk calculated by the triaxial relativistic Hartree-Bogoliubov theory in continuum~(TRHBc) with PC-PK1. For each nucleus, the global and second local minima are respectively represented by the red stars and dots. The energy separation between the contour lines is $0.1$~MeV.}
\label{Fig6}
\end{figure}

To examine the role of triaxial degrees of freedom, in Fig.~\ref{Fig6}, we take the triaxial relativistic Hartree-Bogoliubov theory in continuum~(TRHBc)~\cite{PRC2023KYZhang_TRHB} to study the potential energy surfaces (PESs) in the ($\beta, \gamma$) plane for $^{267}$Bk and $^{331}$Bk, which are predicted to exhibit possible shape coexistence based on the axial DRHBc PECs in Fig.~\ref{Fig5}. The constrained TRHBc calculations are performed with deformation steps of $\Delta\beta=0.05$ and $\Delta\gamma=6^\circ$. Global and local minima are indicated by red stars and dots, respectively. For $^{267}$Bk, the TRHBc theory predicts an oblate ground state at $\beta_2 = -0.15$ and a triaxial local minimum at ($\beta = 0.16, \gamma = 19^\circ$). A clear $\gamma$-soft path connects the two minima, separated by a barrier of approximately $0.3$~MeV. For $^{331}$Bk, the TRHBc calculations yield a nearly prolate ground state at ($\beta = 0.22, \gamma = 4^\circ$) and an oblate local minimum at $\beta_2 = 0.18$, with a barrier height of approximately $0.9$~MeV. These results confirm that the possible shape coexistence predicted in $^{331}$Bk by the DRHBc theory persists when triaxiality is taken into account.

\subsection{Density and rms radius}
\label{subsec.III}
\begin{figure*}
\centering
\includegraphics[width=0.95\linewidth]{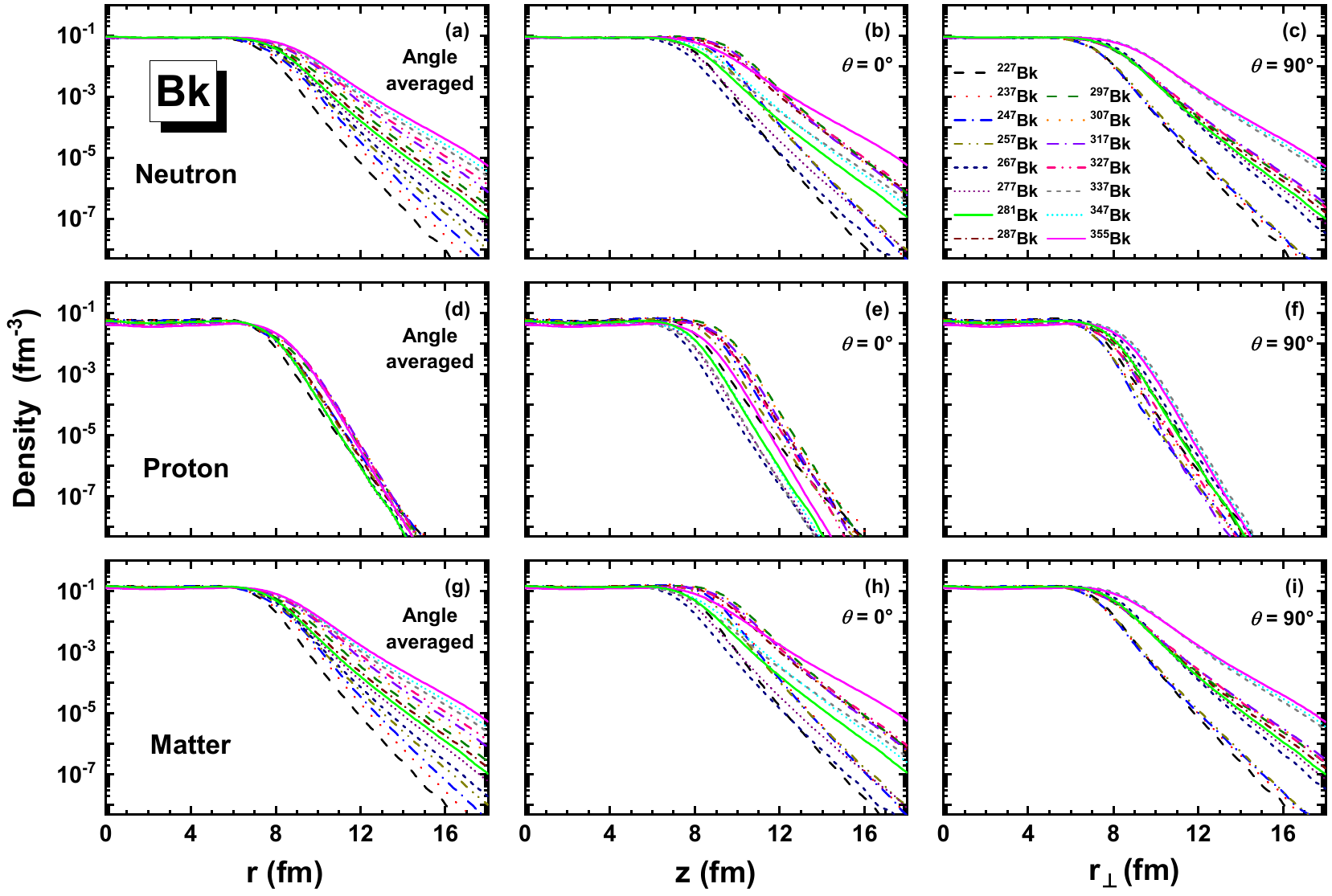}
\caption{(Color online) Density distributions with angle-averaged, along the symmetry axis $z$ ($\theta = 0^\circ$), and perpendicular to the symmetry axis with $r_{\perp} = \sqrt{x^2 + y^2}$ ($\theta = 90^\circ$) are presented for neutrons (panels a, b, c), protons (panels d, e, f), and total nuclear matter (panels g, h, i) of the odd-$A$ Bk isotopes $^{227, 237, 247, ..., 347}$Bk (dashed lines) and $^{281,355}$Bk (solid lines) in the DRHBc calculations with PC-PK1.}
\label{densit}
\end{figure*}

To investigate the evolution of densities, Fig.~\ref{densit} presents the neutron, proton, and matter density distributions for the odd-$A$ Bk isotopes $^{227,237,247,...,347}$Bk. For comparative purposes, the density distributions for $^{281,355}$Bk, with neutron magic numbers $N=184$ and $N=258$, are also plotted and indicated by solid lines. The left, middle, and right panels display the angle-averaged density distributions, the density distributions along the symmetry axis $z$ ($\theta = 0^\circ$), and those perpendicular to the symmetry axis $r_{\perp}$ ($\theta = 90^\circ$), respectively. For the neutron density distributions, as shown in Fig.~\ref{densit}(a), the angle-averaged density distributions generally extend further with increasing neutron number, with the surface expanding outward rapidly while the internal density distribution changes only slightly. By comparing the densities along $\theta = 0^\circ$ and $\theta = 90^\circ$, which reflect deformation effects, it can be observed that the neutron density distribution remains nearly uniform in different directions for the spherical nuclei $^{281,355}$Bk. In contrast, for $^{267,277,337,347}$Bk in oblate shapes, the density is significantly enhanced along $\theta = 90^\circ$, while for Bk isotopes in prolate shapes, it is enhanced along $\theta = 0^\circ$. As a result, the density distributions of the oblate Bk isotopes $^{267,277}$Bk are obviously lower than those of the prolate $^{237}$Bk with much lighter mass along $\theta = 0^\circ$ in the region $z>6$~fm, as shown in Fig.~\ref{densit}(b). Conversely, an opposite trend is observed along $\theta = 90^\circ$, as illustrated in Fig.~\ref{densit}(c).
The density gaps evident in Fig.~\ref{densit}(c) correspond precisely to the shape transition in Bk isotopes from a prolate to an oblate shape. Compared with the neutron density distributions, the proton density distribution exhibits a markedly different and more compact profile. In Fig.~\ref{densit}(d), generally, the inner part of the proton density distribution with $r \leq 6$~fm decreases, while the outer part with $r \geq 7$~fm increases as the neutron number increases. The obvious differences observed in the density distributions along the symmetry axis ($\theta = 0^\circ$) and perpendicular to the symmetry axis ($\theta = 90^\circ$) indicate the presence of deformation in the proton distribution. For the matter density, the evolution pattern is more similar to that of neutrons but is also influenced by the proton density distribution.

\begin{figure}
\centering
\includegraphics[width=0.95\linewidth]{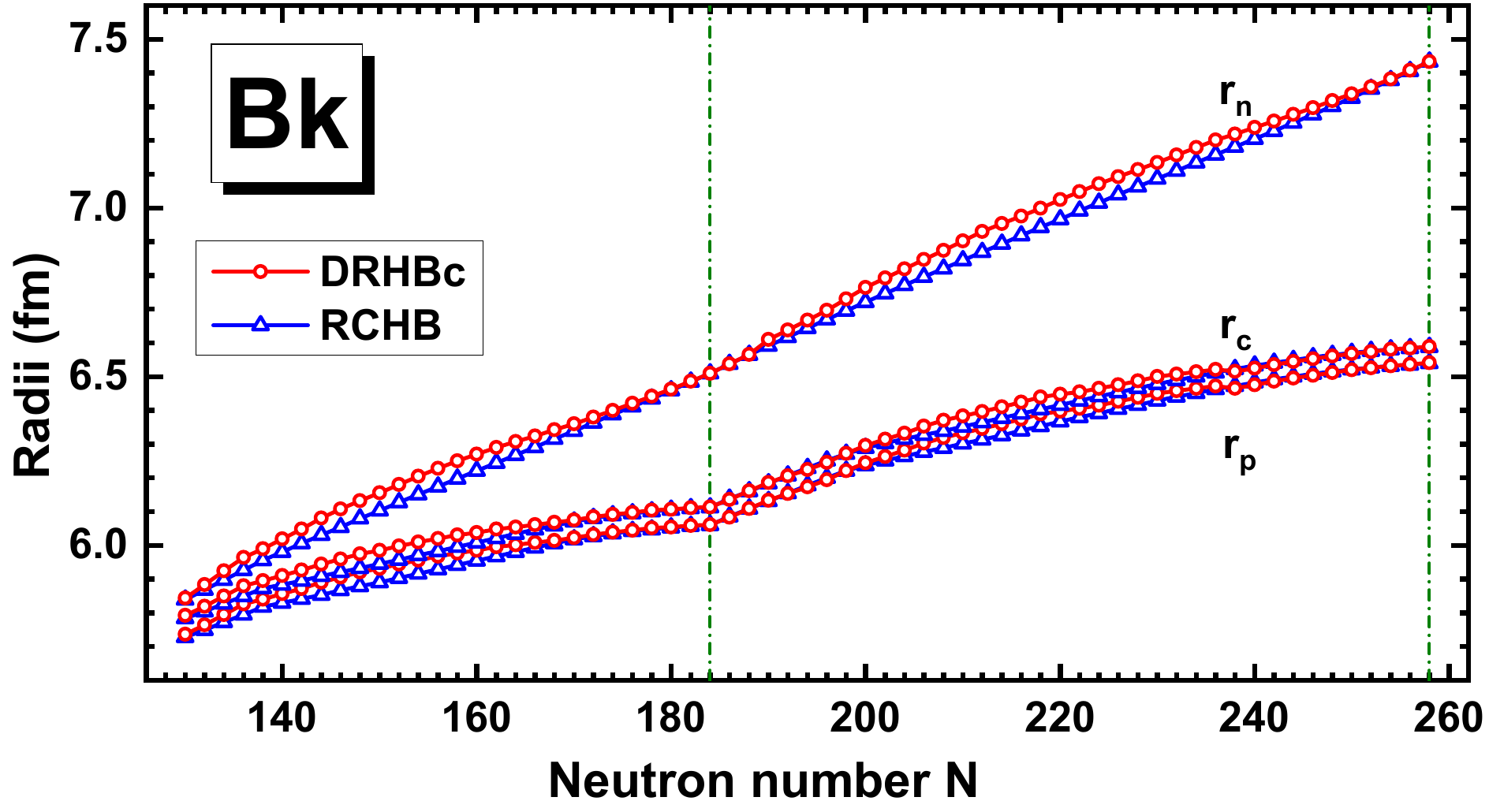}
\caption{(Color online) Rms radii for neutrons $r_{\rm n}$ and protons $r_{\rm p}$, as well as the charge radii $r_{\rm c}$ as functions of the neutron number $N$ in the Bk isotopes obtained by the DRHBc calculations with PC-PK1. The RCHB results~\cite{ADNDT2018Xia_121_1} are also presented for comparison.}
\label{radii}
\end{figure}

Finally, the radii have been examined as a crucial observable that can directly reflect significant features of nuclear structure, such as neutron halos and magic numbers~\cite{PRL2009Nortershauser_102_062503, PRL2008Geithner_101_252502, ADNDT2013Angeli_99_69, ADNDT2021Li_140_101440}. 
The root-mean-square (rms) radius for neutrons and protons is calculated as follows,
\begin{equation}
    r_{\tau} = \langle r^2 \rangle^{1/2} = \sqrt{\frac{\int d^3 \bm{r} [r^2 \rho_{\tau}(\bm{r})]}{N_{\tau}}},
    \label{Eq:radii}
\end{equation}
where $\tau$ denotes neutrons or protons, $\rho_{\tau}(\bm{r})$ represents the corresponding density distribution, and $N_{\tau}$ is the corresponding particle number. The rms charge radius is given by
\begin{equation}
    r_\mathrm{c} = \sqrt{r_{\rm p}^2 + 0.64~\mathrm{fm}^2}.
\end{equation}
In Fig.~\ref{radii}, the neutron, proton, and charge radii of Bk isotopes as functions of neutron number $N$ have been predicted using the DRHBc theory with PC-PK1. 
Distinct kink structures are observed at $N = 184$ and $N = 258$, indicating nuclear shell closures. These observations are consistent with predictions from the two-neutron separation energy $S_{\rm 2n}$, Fermi surface $\lambda_n$, two-neutron gap $\delta_{2n}$, and neutron pairing energy $E_{\rm pair}$. Similar kink structures have also been observed in other mass regions~\cite{ADNDT2022Zhang_144_101488}. Additionally, DRHBc predicted larger radii for the open-shell nuclei compared to those by the spherical RCHB calculations due to the significant deformation effects. Therefore, we conclude that deformation effects have a notable influence on the nuclear radii of open-shell nuclei.

\section[short]{SUMMARY}
\label{sec.IV}

In this work, the ground-state properties and structure evolutions of odd-$A$ transuranium Bk isotopes have been studied by the deformed relativistic Hartree-Bogoliubov theory in continuum, and compared with the available experimental data and the spherical RCHB results. In comparison with RCHB, DRHBc gives a better description for the binding energies $E_{\rm b}$, two-neutron separation energies $S_{\rm 2n}$, two-proton separation energies $S_{\rm 2p}$, and $\alpha$-decay energies $Q_{\alpha}$ for the Bk isotopes, which align well with the experimental data. Significant deformation effects have been verified in the ground-state properties and structures of the Bk isotopes.

First, with the inclusion of deformation in DRHBc, additional binding energies are contributed. Consequently, remarkable crossings of $S_{\rm 2n}$ between RCHB and DRHBc occur within two shell closures, and larger $S_{\rm 2p}$ of the DRHBc are found. This further leads to significantly reduced $Q_{\alpha}$ values before mid-shells by DRHBc. Moreover, crossings of neutron Fermi energy $\lambda_{\rm n}$ between DRHBc and RCHB within two shell closures can also be observed, which can be attributed to the distinct single-particle level structures in the spherical and deformed cases. Meanwhile, the lower proton Fermi energy $\lambda_{\rm p}$ of DRHBc is predicted. Based on $S_{\rm 2n}$ and $S_{\rm 2p}$ along with $\lambda_{\rm n}$ and $\lambda_{\rm p}$, the DRHBc theory predicts that the proton and neutron drip lines of the Bk isotopes are $^{227}$Bk and $^{355}$Bk, respectively.

Second, DRHBc predicts more delicate shell structures. Besides the major shell closures at $N = 184, 258$, which can also be predicted by RCHB, possible sub-shells at $N = 142, 150, 162, 178, 218$, and $230$ are identified by the locations of the peaks of the two-neutron gaps $\delta_{\rm 2n}$ and vanishing neutron pairing energies $E_{\rm pair}^n$ within the framework of DRHBc. Furthermore, distinct kink structures are revealed for $S_{\rm 2p}$ and $\lambda_{\rm p}$ at the shell closures of $N=184, 258$.

Third, according to the evolutions of the ground state deformation $\beta_2$, the nuclear shape undergoes prolate, oblate, and spherical transitions between two shell closures. Obvious prolate dominance has been noted in large mass ranges with $134\leqslant N\leqslant 168$ and $190\leqslant N\leqslant 236$, while nuclear shapes are spherical at shell closures of $N=184, 258$ and oblate in a short range before reaching the shell closures. The spin-parity $K^\pi$ of Bk isotopes in their ground states have been given, with the predominant values $3/2^-$, $7/2^-$, $3/2^+$ and $5/2^+$ in the prolate region while $7/2^-$, $11/2^+$, and $13/2^+$ in the oblate region. Besides, there are some regions with prolate-to-oblate transitions with energy competition between oblate and prolate minima. Possible shape coexistence for $^{267}$Bk and $^{331}$Bk is predicted by DRHBc, which is supported by the TRHBc calculations for $^{331}$Bk but inclined to $\gamma$-soft for $^{267}$Bk.

Finally, based on the DRHBc calculations, the angle-averaged density distributions for neutrons, protons, and nuclear matter, as well as their respective distributions along and perpendicular to the symmetry axis, have been systematically investigated. For $^{281}$Bk and $^{355}$Bk in spherical shapes, the density distributions remain nearly uniform in different directions. In contrast, prolate isotopes exhibit enhanced density distributions along the symmetry axis ($\theta = 0^\circ$), whereas oblate isotopes display increased density along $\theta = 90^\circ$. The evolution of neutron, proton, and charge radii as functions of neutron number has been further examined, revealing pronounced kink structures at neutron shell closures of $N = 184$ and $N = 258$. Compared to spherical RCHB calculations, the inclusion of deformation effects in DRHBc calculations leads to significant enhancements of the radii for open-shell isotopes. 

\begin{acknowledgments}
The authors express their sincere appreciation to members of the DRHBc Mass Table Collaboration for helpful discussions, especially to Dr. Kai-Yuan Zhang, Dr. Cong Pan, and Prof. Jiang-Ming Yao for their valuable insights and meticulous review of the manuscript. This work was partly supported by the Natural Science Foundation of Henan Province (No.~242300421156), the National Natural Science Foundation of China (No.~U2032141, No.~12481540215, and No.~12435006), the Open Project of Guangxi Key Laboratory of Nuclear Physics and Nuclear Technology (No.~NLK2022-02), the Central Government Guidance Funds for Local Scientific and Technological Development, China (No.~Guike ZY22096024), National Key R\&D Program of China (No.~2024YFE0109803), State Key Laboratory of Nuclear Physics and Technology, Peking University (No.~NPT2023ZX03), and the Fundamental Research Funds for the Central Universities.
\end{acknowledgments}


\begin{thebibliography}{103}
\expandafter\ifx\csname natexlab\endcsname\relax\def\natexlab#1{#1}\fi
\expandafter\ifx\csname bibnamefont\endcsname\relax
  \def\bibnamefont#1{#1}\fi
\expandafter\ifx\csname bibfnamefont\endcsname\relax
  \def\bibfnamefont#1{#1}\fi
\expandafter\ifx\csname citenamefont\endcsname\relax
  \def\citenamefont#1{#1}\fi
\expandafter\ifx\csname url\endcsname\relax
  \def\url#1{\texttt{#1}}\fi
\expandafter\ifx\csname urlprefix\endcsname\relax\def\urlprefix{URL }\fi
\providecommand{\bibinfo}[2]{#2}
\providecommand{\eprint}[2][]{\url{#2}}

\bibitem[{\citenamefont{Thoennessen}(2016)}]{Thoennessen2016}
\bibinfo{author}{\bibfnamefont{M.}~\bibnamefont{Thoennessen}},
  \emph{\bibinfo{title}{Transuranium Elements}} (\bibinfo{publisher}{Springer},
  \bibinfo{address}{New York}, \bibinfo{year}{2016}).

\bibitem[{\citenamefont{Morita et~al.}(2004)\citenamefont{Morita, Morimoto,
  Kaji et~al.}}]{JPSJ2004Morita_73_2593}
\bibinfo{author}{\bibfnamefont{K.}~\bibnamefont{Morita}},
  \bibinfo{author}{\bibfnamefont{K.}~\bibnamefont{Morimoto}},
  \bibinfo{author}{\bibfnamefont{D.}~\bibnamefont{Kaji}}, \bibnamefont{et~al.},
  \bibinfo{journal}{J. Phys. Soc. Jpn.} \textbf{\bibinfo{volume}{73}},
  \bibinfo{pages}{2593} (\bibinfo{year}{2004}).

\bibitem[{\citenamefont{Cai}(2006)}]{Cai2006}
\bibinfo{author}{\bibfnamefont{S.~Y.} \bibnamefont{Cai}},
  \emph{\bibinfo{title}{Man-made elements(in Chinese)}}
  (\bibinfo{publisher}{Shanghai Popular Science Press},
  \bibinfo{address}{Shanghai}, \bibinfo{year}{2006}).

\bibitem[{\citenamefont{Myers and Swiatecki}(1966)}]{NP1966Myers_81_1}
\bibinfo{author}{\bibfnamefont{W.~D.} \bibnamefont{Myers}} \bibnamefont{and}
  \bibinfo{author}{\bibfnamefont{W.~J.} \bibnamefont{Swiatecki}},
  \bibinfo{journal}{Nucl. Phys.} \textbf{\bibinfo{volume}{81}},
  \bibinfo{pages}{1} (\bibinfo{year}{1966}).

\bibitem[{\citenamefont{Wong}(1966)}]{PL1966Wong_21_688}
\bibinfo{author}{\bibfnamefont{C.~Y.} \bibnamefont{Wong}},
  \bibinfo{journal}{Phys. Lett.} \textbf{\bibinfo{volume}{21}},
  \bibinfo{pages}{688} (\bibinfo{year}{1966}).

\bibitem[{\citenamefont{Sobiczewski et~al.}(1966)\citenamefont{Sobiczewski,
  Gareev, and Kalinkin}}]{PL1966Sobiczewski_22_500}
\bibinfo{author}{\bibfnamefont{A.}~\bibnamefont{Sobiczewski}},
  \bibinfo{author}{\bibfnamefont{F.~A.} \bibnamefont{Gareev}},
  \bibnamefont{and} \bibinfo{author}{\bibfnamefont{B.~N.}
  \bibnamefont{Kalinkin}}, \bibinfo{journal}{Phys. Lett.}
  \textbf{\bibinfo{volume}{22}}, \bibinfo{pages}{500} (\bibinfo{year}{1966}).

\bibitem[{\citenamefont{Sobiczewski et~al.}(1969)\citenamefont{Sobiczewski,
  Szymański, Wycech et~al.}}]{NPA1969Sobiczewski_131_67}
\bibinfo{author}{\bibfnamefont{A.}~\bibnamefont{Sobiczewski}},
  \bibinfo{author}{\bibfnamefont{Z.}~\bibnamefont{Szymański}},
  \bibinfo{author}{\bibfnamefont{S.}~\bibnamefont{Wycech}},
  \bibnamefont{et~al.}, \bibinfo{journal}{Nucl. Phys. A}
  \textbf{\bibinfo{volume}{131}}, \bibinfo{pages}{67} (\bibinfo{year}{1969}).

\bibitem[{\citenamefont{Mosel and Greiner}(1969)}]{ZAP1969Mosel_222_261}
\bibinfo{author}{\bibfnamefont{U.}~\bibnamefont{Mosel}} \bibnamefont{and}
  \bibinfo{author}{\bibfnamefont{W.}~\bibnamefont{Greiner}},
  \bibinfo{journal}{Z. Angew. Phys.} \textbf{\bibinfo{volume}{222}},
  \bibinfo{pages}{261} (\bibinfo{year}{1969}).

\bibitem[{\citenamefont{Zhou}(2017)}]{NPR2017Zhou_34_318}
\bibinfo{author}{\bibfnamefont{S.~G.} \bibnamefont{Zhou}},
  \bibinfo{journal}{Nucl. Phys. Rev.} \textbf{\bibinfo{volume}{34}},
  \bibinfo{pages}{318} (\bibinfo{year}{2017}).

\bibitem[{\citenamefont{Zhang et~al.}(2022{\natexlab{a}})\citenamefont{Zhang,
  Zhang, Zhang, Tang, Chang, Li, and Cheng}}]{JBNU2022Zhang_58_392}
\bibinfo{author}{\bibfnamefont{F.~S.} \bibnamefont{Zhang}},
  \bibinfo{author}{\bibfnamefont{Y.~H.} \bibnamefont{Zhang}},
  \bibinfo{author}{\bibfnamefont{M.~H.} \bibnamefont{Zhang}},
  \bibinfo{author}{\bibfnamefont{N.}~\bibnamefont{Tang}},
  \bibinfo{author}{\bibfnamefont{S.~H.} \bibnamefont{Chang}},
  \bibinfo{author}{\bibfnamefont{J.~J.} \bibnamefont{Li}}, \bibnamefont{and}
  \bibinfo{author}{\bibfnamefont{W.}~\bibnamefont{Cheng}}, \bibinfo{journal}{J.
  Beijing Norm. Univ.} \textbf{\bibinfo{volume}{58}}, \bibinfo{pages}{392}
  (\bibinfo{year}{2022}{\natexlab{a}}).

\bibitem[{\citenamefont{Yang et~al.}(2020)\citenamefont{Yang, Zeng, Xiao
  et~al.}}]{CSB2020Yang_65_8}
\bibinfo{author}{\bibfnamefont{J.~C.} \bibnamefont{Yang}},
  \bibinfo{author}{\bibfnamefont{G.}~\bibnamefont{Zeng}},
  \bibinfo{author}{\bibfnamefont{G.~Q.} \bibnamefont{Xiao}},
  \bibnamefont{et~al.}, \bibinfo{journal}{Chin. Sci. Bull.}
  \textbf{\bibinfo{volume}{65}}, \bibinfo{pages}{8} (\bibinfo{year}{2020}).

\bibitem[{\citenamefont{Seife}(2005)}]{Science2005Seife_309_78}
\bibinfo{author}{\bibfnamefont{C.}~\bibnamefont{Seife}},
  \bibinfo{journal}{Science} \textbf{\bibinfo{volume}{309}},
  \bibinfo{pages}{78} (\bibinfo{year}{2005}).

\bibitem[{\citenamefont{Zhou}(2014)}]{PHYSICS2014Zhou_43_817}
\bibinfo{author}{\bibfnamefont{S.~G.} \bibnamefont{Zhou}},
  \bibinfo{journal}{Physics} \textbf{\bibinfo{volume}{43}},
  \bibinfo{pages}{817} (\bibinfo{year}{2014}).

\bibitem[{\citenamefont{Li et~al.}(2014{\natexlab{a}})\citenamefont{Li, L{\"u},
  Wang, Wen, Xia, Zhang, Zhao, Zhao, and Zhou}}]{NPR2014Li_31_253}
\bibinfo{author}{\bibfnamefont{L.~L.} \bibnamefont{Li}},
  \bibinfo{author}{\bibfnamefont{B.~N.} \bibnamefont{L{\"u}}},
  \bibinfo{author}{\bibfnamefont{N.}~\bibnamefont{Wang}},
  \bibinfo{author}{\bibfnamefont{K.}~\bibnamefont{Wen}},
  \bibinfo{author}{\bibfnamefont{C.~J.} \bibnamefont{Xia}},
  \bibinfo{author}{\bibfnamefont{Z.~H.} \bibnamefont{Zhang}},
  \bibinfo{author}{\bibfnamefont{J.}~\bibnamefont{Zhao}},
  \bibinfo{author}{\bibfnamefont{E.~G.} \bibnamefont{Zhao}}, \bibnamefont{and}
  \bibinfo{author}{\bibfnamefont{S.~G.} \bibnamefont{Zhou}},
  \bibinfo{journal}{Nucl. Phys. Rev.} \textbf{\bibinfo{volume}{31}},
  \bibinfo{pages}{253} (\bibinfo{year}{2014}{\natexlab{a}}).

\bibitem[{\citenamefont{L{\"u} et~al.}(2016)\citenamefont{L{\"u}, Zhao, Zhao,
  and Zhou}}]{Lv2016Superheavy}
\bibinfo{author}{\bibfnamefont{B.~N.} \bibnamefont{L{\"u}}},
  \bibinfo{author}{\bibfnamefont{J.}~\bibnamefont{Zhao}},
  \bibinfo{author}{\bibfnamefont{E.~G.} \bibnamefont{Zhao}}, \bibnamefont{and}
  \bibinfo{author}{\bibfnamefont{S.~G.} \bibnamefont{Zhou}},
  \emph{\bibinfo{title}{Relativistic Density Functional for Nuclear Structure}}
  (\bibinfo{publisher}{World Scientific}, \bibinfo{year}{2016}), chap.
  \bibinfo{chapter}{5 Superheavy nuclei and fission barriers}, pp.
  \bibinfo{pages}{171--217}.

\bibitem[{\citenamefont{He et~al.}(2024)\citenamefont{He, Wu, Zhang, and
  Shen}}]{PRC2024He_110_014301}
\bibinfo{author}{\bibfnamefont{X.~T.} \bibnamefont{He}},
  \bibinfo{author}{\bibfnamefont{J.~W.} \bibnamefont{Wu}},
  \bibinfo{author}{\bibfnamefont{X.~Y.} \bibnamefont{Zhang}}, \bibnamefont{and}
  \bibinfo{author}{\bibfnamefont{C.~W.} \bibnamefont{Shen}},
  \bibinfo{journal}{Phys. Rev. C} \textbf{\bibinfo{volume}{110}},
  \bibinfo{pages}{104301} (\bibinfo{year}{2024}).

\bibitem[{\citenamefont{Ismail et~al.}(2024)\citenamefont{Ismail, Abd-Elnasser,
  Adel, Abdul-Magead, and Elsharkawy}}]{PRC2024Ismail_109_014606}
\bibinfo{author}{\bibfnamefont{M.}~\bibnamefont{Ismail}},
  \bibinfo{author}{\bibfnamefont{S.~G.} \bibnamefont{Abd-Elnasser}},
  \bibinfo{author}{\bibfnamefont{A.}~\bibnamefont{Adel}},
  \bibinfo{author}{\bibfnamefont{I.~A.~M.} \bibnamefont{Abdul-Magead}},
  \bibnamefont{and} \bibinfo{author}{\bibfnamefont{H.~M.}
  \bibnamefont{Elsharkawy}}, \bibinfo{journal}{Phys. Rev. C}
  \textbf{\bibinfo{volume}{109}}, \bibinfo{pages}{014606}
  (\bibinfo{year}{2024}).

\bibitem[{\citenamefont{Denisov}(2024)}]{PRC2024Denisov_109_044618}
\bibinfo{author}{\bibfnamefont{V.~Y.} \bibnamefont{Denisov}},
  \bibinfo{journal}{Phys. Rev. C} \textbf{\bibinfo{volume}{109}},
  \bibinfo{pages}{044618} (\bibinfo{year}{2024}).

\bibitem[{\citenamefont{Zhang et~al.}(2013)\citenamefont{Zhang, Wen, He, Zeng,
  Zhao, and Zhou}}]{NPR2013Zhang_30_268}
\bibinfo{author}{\bibfnamefont{Z.~H.} \bibnamefont{Zhang}},
  \bibinfo{author}{\bibfnamefont{K.}~\bibnamefont{Wen}},
  \bibinfo{author}{\bibfnamefont{X.~T.} \bibnamefont{He}},
  \bibinfo{author}{\bibfnamefont{J.~Y.} \bibnamefont{Zeng}},
  \bibinfo{author}{\bibfnamefont{E.~G.} \bibnamefont{Zhao}}, \bibnamefont{and}
  \bibinfo{author}{\bibfnamefont{S.~G.} \bibnamefont{Zhou}},
  \bibinfo{journal}{Nucl. Phys. Rev.} \textbf{\bibinfo{volume}{30}},
  \bibinfo{pages}{268} (\bibinfo{year}{2013}).

\bibitem[{\citenamefont{Wen et~al.}(2012)\citenamefont{Wen, Zhang, Zhao, and
  Zhou}}]{SS2012Wen_42_22}
\bibinfo{author}{\bibfnamefont{K.}~\bibnamefont{Wen}},
  \bibinfo{author}{\bibfnamefont{Z.~H.} \bibnamefont{Zhang}},
  \bibinfo{author}{\bibfnamefont{E.~G.} \bibnamefont{Zhao}}, \bibnamefont{and}
  \bibinfo{author}{\bibfnamefont{S.~G.} \bibnamefont{Zhou}},
  \bibinfo{journal}{Sci. Sin.} \textbf{\bibinfo{volume}{42}},
  \bibinfo{pages}{22} (\bibinfo{year}{2012}).

\bibitem[{\citenamefont{{\'C}wiok et~al.}(1994)\citenamefont{{\'C}wiok,
  Hofmann, and Nazarewicz}}]{NPA1994Cwiok_573_356}
\bibinfo{author}{\bibfnamefont{S.}~\bibnamefont{{\'C}wiok}},
  \bibinfo{author}{\bibfnamefont{S.}~\bibnamefont{Hofmann}}, \bibnamefont{and}
  \bibinfo{author}{\bibfnamefont{W.}~\bibnamefont{Nazarewicz}},
  \bibinfo{journal}{Nucl. Phys. A} \textbf{\bibinfo{volume}{573}},
  \bibinfo{pages}{356} (\bibinfo{year}{1994}).

\bibitem[{\citenamefont{Adamian et~al.}(2011)\citenamefont{Adamian, Antonenko,
  Kuklin, L{\"u}, Malov, and Zhou}}]{PRC2011Adamian_84_024324}
\bibinfo{author}{\bibfnamefont{G.~G.} \bibnamefont{Adamian}},
  \bibinfo{author}{\bibfnamefont{N.~V.} \bibnamefont{Antonenko}},
  \bibinfo{author}{\bibfnamefont{S.~N.} \bibnamefont{Kuklin}},
  \bibinfo{author}{\bibfnamefont{B.~N.} \bibnamefont{L{\"u}}},
  \bibinfo{author}{\bibfnamefont{L.~A.} \bibnamefont{Malov}}, \bibnamefont{and}
  \bibinfo{author}{\bibfnamefont{S.~G.} \bibnamefont{Zhou}},
  \bibinfo{journal}{Phys. Rev. C} \textbf{\bibinfo{volume}{84}},
  \bibinfo{pages}{024324} (\bibinfo{year}{2011}).

\bibitem[{\citenamefont{Xu et~al.}(2004)\citenamefont{Xu, Zhao, Wyss, and
  Walker}}]{PRL2004Xu_92_252501}
\bibinfo{author}{\bibfnamefont{F.~R.} \bibnamefont{Xu}},
  \bibinfo{author}{\bibfnamefont{E.~G.} \bibnamefont{Zhao}},
  \bibinfo{author}{\bibfnamefont{R.}~\bibnamefont{Wyss}}, \bibnamefont{and}
  \bibinfo{author}{\bibfnamefont{P.~M.} \bibnamefont{Walker}},
  \bibinfo{journal}{Phys. Rev. Lett.} \textbf{\bibinfo{volume}{92}},
  \bibinfo{pages}{252501} (\bibinfo{year}{2004}).

\bibitem[{\citenamefont{Liu et~al.}(2012)\citenamefont{Liu, Xu, and
  Walker}}]{PRC2012Liu_86_011301}
\bibinfo{author}{\bibfnamefont{H.~L.} \bibnamefont{Liu}},
  \bibinfo{author}{\bibfnamefont{F.~R.} \bibnamefont{Xu}}, \bibnamefont{and}
  \bibinfo{author}{\bibfnamefont{P.~M.} \bibnamefont{Walker}},
  \bibinfo{journal}{Phys. Rev. C} \textbf{\bibinfo{volume}{86}},
  \bibinfo{pages}{011301} (\bibinfo{year}{2012}).

\bibitem[{\citenamefont{Bender et~al.}(1998)\citenamefont{Bender, Rutz,
  Reinhard, Maruhn, and Greiner}}]{PRC1998Bender_58_2126}
\bibinfo{author}{\bibfnamefont{M.}~\bibnamefont{Bender}},
  \bibinfo{author}{\bibfnamefont{K.}~\bibnamefont{Rutz}},
  \bibinfo{author}{\bibfnamefont{P.-G.} \bibnamefont{Reinhard}},
  \bibinfo{author}{\bibfnamefont{J.~A.} \bibnamefont{Maruhn}},
  \bibnamefont{and} \bibinfo{author}{\bibfnamefont{W.}~\bibnamefont{Greiner}},
  \bibinfo{journal}{Phys. Rev. C} \textbf{\bibinfo{volume}{58}},
  \bibinfo{pages}{2126} (\bibinfo{year}{1998}).

\bibitem[{\citenamefont{Delaroche et~al.}(2006)\citenamefont{Delaroche, Girod,
  Goutte, and Libert}}]{NPA2006Delaroche_771_103}
\bibinfo{author}{\bibfnamefont{J.~P.} \bibnamefont{Delaroche}},
  \bibinfo{author}{\bibfnamefont{M.}~\bibnamefont{Girod}},
  \bibinfo{author}{\bibfnamefont{H.}~\bibnamefont{Goutte}}, \bibnamefont{and}
  \bibinfo{author}{\bibfnamefont{J.}~\bibnamefont{Libert}},
  \bibinfo{journal}{Nucl. Phys. A} \textbf{\bibinfo{volume}{771}},
  \bibinfo{pages}{103} (\bibinfo{year}{2006}).

\bibitem[{\citenamefont{Jolos et~al.}(2011)\citenamefont{Jolos, Malov,
  Shirikova, and Sushkov}}]{JPG2011Jolos_38_115103}
\bibinfo{author}{\bibfnamefont{R.~V.} \bibnamefont{Jolos}},
  \bibinfo{author}{\bibfnamefont{L.~A.} \bibnamefont{Malov}},
  \bibinfo{author}{\bibfnamefont{N.~Y.} \bibnamefont{Shirikova}},
  \bibnamefont{and} \bibinfo{author}{\bibfnamefont{A.~V.}
  \bibnamefont{Sushkov}}, \bibinfo{journal}{J. Phys. G: Nucl. Part. Phys.}
  \textbf{\bibinfo{volume}{38}}, \bibinfo{pages}{115103}
  (\bibinfo{year}{2011}).

\bibitem[{\citenamefont{Zhuang et~al.}(2012)\citenamefont{Zhuang, Li, and
  Liu}}]{CTP2012Zhuang_57_271}
\bibinfo{author}{\bibfnamefont{K.}~\bibnamefont{Zhuang}},
  \bibinfo{author}{\bibfnamefont{Z.~B.} \bibnamefont{Li}}, \bibnamefont{and}
  \bibinfo{author}{\bibfnamefont{Y.~X.} \bibnamefont{Liu}},
  \bibinfo{journal}{Commun. Theor. Phys.} \textbf{\bibinfo{volume}{57}},
  \bibinfo{pages}{271} (\bibinfo{year}{2012}).

\bibitem[{\citenamefont{Sun et~al.}(2008)\citenamefont{Sun, Long, Al-Khudair,
  and Sheikh}}]{PRC2008Sun_77_044307}
\bibinfo{author}{\bibfnamefont{Y.}~\bibnamefont{Sun}},
  \bibinfo{author}{\bibfnamefont{G.~L.} \bibnamefont{Long}},
  \bibinfo{author}{\bibfnamefont{F.}~\bibnamefont{Al-Khudair}},
  \bibnamefont{and} \bibinfo{author}{\bibfnamefont{J.~A.}
  \bibnamefont{Sheikh}}, \bibinfo{journal}{Phys. Rev. C}
  \textbf{\bibinfo{volume}{77}}, \bibinfo{pages}{044307}
  (\bibinfo{year}{2008}).

\bibitem[{\citenamefont{Chen et~al.}(2008)\citenamefont{Chen, Sun, and
  Gao}}]{PRC2008Chen_77_061305}
\bibinfo{author}{\bibfnamefont{Y.~S.} \bibnamefont{Chen}},
  \bibinfo{author}{\bibfnamefont{Y.}~\bibnamefont{Sun}}, \bibnamefont{and}
  \bibinfo{author}{\bibfnamefont{Z.~C.} \bibnamefont{Gao}},
  \bibinfo{journal}{Phys. Rev. C} \textbf{\bibinfo{volume}{77}},
  \bibinfo{pages}{061305} (\bibinfo{year}{2008}).

\bibitem[{\citenamefont{Al-Khudair et~al.}(2009)\citenamefont{Al-Khudair, Long,
  and Sun}}]{PRC2009Al-Khudair_79_034320}
\bibinfo{author}{\bibfnamefont{F.}~\bibnamefont{Al-Khudair}},
  \bibinfo{author}{\bibfnamefont{G.~L.} \bibnamefont{Long}}, \bibnamefont{and}
  \bibinfo{author}{\bibfnamefont{Y.}~\bibnamefont{Sun}},
  \bibinfo{journal}{Phys. Rev. C} \textbf{\bibinfo{volume}{79}},
  \bibinfo{pages}{034320} (\bibinfo{year}{2009}).

\bibitem[{\citenamefont{Cui et~al.}(2014)\citenamefont{Cui, Zhou, Chen, Sun,
  Wu, and Gao}}]{PRC2014Cui_90_014321}
\bibinfo{author}{\bibfnamefont{J.~W.} \bibnamefont{Cui}},
  \bibinfo{author}{\bibfnamefont{X.~R.} \bibnamefont{Zhou}},
  \bibinfo{author}{\bibfnamefont{F.~Q.} \bibnamefont{Chen}},
  \bibinfo{author}{\bibfnamefont{Y.}~\bibnamefont{Sun}},
  \bibinfo{author}{\bibfnamefont{C.~L.} \bibnamefont{Wu}}, \bibnamefont{and}
  \bibinfo{author}{\bibfnamefont{Z.~C.} \bibnamefont{Gao}},
  \bibinfo{journal}{Phys. Rev. C} \textbf{\bibinfo{volume}{90}},
  \bibinfo{pages}{014321} (\bibinfo{year}{2014}).

\bibitem[{\citenamefont{Egido and Ring}(1982)}]{JPG1982Egido_8_L43}
\bibinfo{author}{\bibfnamefont{J.}~\bibnamefont{Egido}} \bibnamefont{and}
  \bibinfo{author}{\bibfnamefont{P.}~\bibnamefont{Ring}}, \bibinfo{journal}{J.
  Phys. G: Nucl. Phys.} \textbf{\bibinfo{volume}{8}}, \bibinfo{pages}{L43}
  (\bibinfo{year}{1982}).

\bibitem[{\citenamefont{Egido and Ring}(1984)}]{NPA1984Egido_423_93}
\bibinfo{author}{\bibfnamefont{J.~L.} \bibnamefont{Egido}} \bibnamefont{and}
  \bibinfo{author}{\bibfnamefont{P.}~\bibnamefont{Ring}},
  \bibinfo{journal}{Nucl. Phys. A} \textbf{\bibinfo{volume}{423}},
  \bibinfo{pages}{93} (\bibinfo{year}{1984}).

\bibitem[{\citenamefont{He et~al.}(2009)\citenamefont{He, Ren, Liu, and
  Zhao}}]{NPA2009He_817_45}
\bibinfo{author}{\bibfnamefont{X.~T.} \bibnamefont{He}},
  \bibinfo{author}{\bibfnamefont{Z.~Z.} \bibnamefont{Ren}},
  \bibinfo{author}{\bibfnamefont{S.~X.} \bibnamefont{Liu}}, \bibnamefont{and}
  \bibinfo{author}{\bibfnamefont{E.~G.} \bibnamefont{Zhao}},
  \bibinfo{journal}{Nucl. Phys. A} \textbf{\bibinfo{volume}{817}},
  \bibinfo{pages}{45} (\bibinfo{year}{2009}).

\bibitem[{\citenamefont{Mo et~al.}(2014)\citenamefont{Mo, Liu, and
  Wang}}]{PRC2014Mo_90_024320}
\bibinfo{author}{\bibfnamefont{Q.~H.} \bibnamefont{Mo}},
  \bibinfo{author}{\bibfnamefont{M.}~\bibnamefont{Liu}}, \bibnamefont{and}
  \bibinfo{author}{\bibfnamefont{N.}~\bibnamefont{Wang}},
  \bibinfo{journal}{Phys. Rev. C} \textbf{\bibinfo{volume}{90}},
  \bibinfo{pages}{024320} (\bibinfo{year}{2014}).

\bibitem[{\citenamefont{Rutz et~al.}(1997)\citenamefont{Rutz, Bender,
  B{\"u}rvenich et~al.}}]{PRC1997Rutz_56_238}
\bibinfo{author}{\bibfnamefont{K.}~\bibnamefont{Rutz}},
  \bibinfo{author}{\bibfnamefont{M.}~\bibnamefont{Bender}},
  \bibinfo{author}{\bibfnamefont{T.}~\bibnamefont{B{\"u}rvenich}},
  \bibnamefont{et~al.}, \bibinfo{journal}{Phys. Rev. C}
  \textbf{\bibinfo{volume}{56}}, \bibinfo{pages}{238} (\bibinfo{year}{1997}).

\bibitem[{\citenamefont{Zhang et~al.}(2005)\citenamefont{Zhang, Meng, Zhang,
  Geng, and Toki}}]{NPA2005Zhang_753_106}
\bibinfo{author}{\bibfnamefont{W.}~\bibnamefont{Zhang}},
  \bibinfo{author}{\bibfnamefont{J.}~\bibnamefont{Meng}},
  \bibinfo{author}{\bibfnamefont{S.}~\bibnamefont{Zhang}},
  \bibinfo{author}{\bibfnamefont{L.}~\bibnamefont{Geng}}, \bibnamefont{and}
  \bibinfo{author}{\bibfnamefont{H.}~\bibnamefont{Toki}},
  \bibinfo{journal}{Nucl. Phys. A} \textbf{\bibinfo{volume}{753}},
  \bibinfo{pages}{106} (\bibinfo{year}{2005}).

\bibitem[{\citenamefont{Zhou and Sagawa}(2012)}]{JPG2012Zhou_39_085104}
\bibinfo{author}{\bibfnamefont{X.~R.} \bibnamefont{Zhou}} \bibnamefont{and}
  \bibinfo{author}{\bibfnamefont{H.}~\bibnamefont{Sagawa}},
  \bibinfo{journal}{J. Phys. G: Nucl. Part. Phys.}
  \textbf{\bibinfo{volume}{39}}, \bibinfo{pages}{085104}
  (\bibinfo{year}{2012}).

\bibitem[{\citenamefont{Li et~al.}(2014{\natexlab{b}})\citenamefont{Li, Long,
  Margueron, and van Giai}}]{PLB2014li_732_169}
\bibinfo{author}{\bibfnamefont{J.~J.} \bibnamefont{Li}},
  \bibinfo{author}{\bibfnamefont{W.~H.} \bibnamefont{Long}},
  \bibinfo{author}{\bibfnamefont{J.}~\bibnamefont{Margueron}},
  \bibnamefont{and} \bibinfo{author}{\bibfnamefont{N.}~\bibnamefont{van Giai}},
  \bibinfo{journal}{Phys. Lett. B} \textbf{\bibinfo{volume}{732}},
  \bibinfo{pages}{169} (\bibinfo{year}{2014}{\natexlab{b}}).

\bibitem[{\citenamefont{Patyk and Sobiczewski}(1991)}]{NPA1991Patyk_533_132}
\bibinfo{author}{\bibfnamefont{Z.}~\bibnamefont{Patyk}} \bibnamefont{and}
  \bibinfo{author}{\bibfnamefont{A.}~\bibnamefont{Sobiczewski}},
  \bibinfo{journal}{Nucl. Phys. A} \textbf{\bibinfo{volume}{533}},
  \bibinfo{pages}{132} (\bibinfo{year}{1991}).

\bibitem[{\citenamefont{Pei et~al.}(2003)\citenamefont{Pei, Xu, Wu, and
  Zhao}}]{NPR2003Pei_20_116}
\bibinfo{author}{\bibfnamefont{J.~C.} \bibnamefont{Pei}},
  \bibinfo{author}{\bibfnamefont{F.~R.} \bibnamefont{Xu}},
  \bibinfo{author}{\bibfnamefont{Z.~Y.} \bibnamefont{Wu}}, \bibnamefont{and}
  \bibinfo{author}{\bibfnamefont{E.~G.} \bibnamefont{Zhao}},
  \bibinfo{journal}{Nucl. Phys. Rev.} \textbf{\bibinfo{volume}{20}},
  \bibinfo{pages}{116} (\bibinfo{year}{2003}).

\bibitem[{\citenamefont{Bohr and
  Mottelson}(1998)}]{BOOK1998Bohr_NuclearStructure}
\bibinfo{author}{\bibfnamefont{A.}~\bibnamefont{Bohr}} \bibnamefont{and}
  \bibinfo{author}{\bibfnamefont{B.~R.} \bibnamefont{Mottelson}},
  \emph{\bibinfo{title}{Nuclear Structure}} (\bibinfo{publisher}{World
  Scientific}, \bibinfo{address}{New York}, \bibinfo{year}{1998}).

\bibitem[{\citenamefont{Sun et~al.}(2024)\citenamefont{Sun, Li, and
  Liu}}]{PRC2024SunTT}
\bibinfo{author}{\bibfnamefont{T.~T.} \bibnamefont{Sun}},
  \bibinfo{author}{\bibfnamefont{B.~X.} \bibnamefont{Li}}, \bibnamefont{and}
  \bibinfo{author}{\bibfnamefont{K.}~\bibnamefont{Liu}},
  \bibinfo{journal}{Phys. Rev. C} \textbf{\bibinfo{volume}{109}},
  \bibinfo{pages}{014323} (\bibinfo{year}{2024}).

\bibitem[{\citenamefont{Zhou et~al.}(2010)\citenamefont{Zhou, Meng, Ring, and
  Zhao}}]{PRC2010Zhou_82_011301}
\bibinfo{author}{\bibfnamefont{S.~G.} \bibnamefont{Zhou}},
  \bibinfo{author}{\bibfnamefont{J.}~\bibnamefont{Meng}},
  \bibinfo{author}{\bibfnamefont{P.}~\bibnamefont{Ring}}, \bibnamefont{and}
  \bibinfo{author}{\bibfnamefont{E.~G.} \bibnamefont{Zhao}},
  \bibinfo{journal}{Phys. Rev. C} \textbf{\bibinfo{volume}{82}},
  \bibinfo{pages}{011301} (\bibinfo{year}{2010}).

\bibitem[{\citenamefont{Zhou et~al.}(2011)\citenamefont{Zhou, Meng, Ring, and
  Zhao}}]{JP2011Zhou_312_092067}
\bibinfo{author}{\bibfnamefont{S.~G.} \bibnamefont{Zhou}},
  \bibinfo{author}{\bibfnamefont{J.}~\bibnamefont{Meng}},
  \bibinfo{author}{\bibfnamefont{P.}~\bibnamefont{Ring}}, \bibnamefont{and}
  \bibinfo{author}{\bibfnamefont{E.~G.} \bibnamefont{Zhao}},
  \bibinfo{journal}{J. Phys.} \textbf{\bibinfo{volume}{312}},
  \bibinfo{pages}{092067} (\bibinfo{year}{2011}).

\bibitem[{\citenamefont{Li et~al.}(2012{\natexlab{a}})\citenamefont{Li, Meng,
  Ring, Zhao, and Zhou}}]{AIP2012Zhou_1491_208}
\bibinfo{author}{\bibfnamefont{L.~L.} \bibnamefont{Li}},
  \bibinfo{author}{\bibfnamefont{J.}~\bibnamefont{Meng}},
  \bibinfo{author}{\bibfnamefont{P.}~\bibnamefont{Ring}},
  \bibinfo{author}{\bibfnamefont{E.~G.} \bibnamefont{Zhao}}, \bibnamefont{and}
  \bibinfo{author}{\bibfnamefont{S.~G.} \bibnamefont{Zhou}},
  \bibinfo{journal}{AIP Conference Proceedings}
  \textbf{\bibinfo{volume}{1491}}, \bibinfo{pages}{208}
  (\bibinfo{year}{2012}{\natexlab{a}}).

\bibitem[{\citenamefont{Sun et~al.}(2020{\natexlab{a}})\citenamefont{Sun, Qian,
  Chen, Ring, and Li}}]{PRC2020Sun_101_014321}
\bibinfo{author}{\bibfnamefont{T.~T.} \bibnamefont{Sun}},
  \bibinfo{author}{\bibfnamefont{L.}~\bibnamefont{Qian}},
  \bibinfo{author}{\bibfnamefont{C.}~\bibnamefont{Chen}},
  \bibinfo{author}{\bibfnamefont{P.}~\bibnamefont{Ring}}, \bibnamefont{and}
  \bibinfo{author}{\bibfnamefont{Z.~P.} \bibnamefont{Li}},
  \bibinfo{journal}{Phys. Rev. C} \textbf{\bibinfo{volume}{101}},
  \bibinfo{pages}{014321} (\bibinfo{year}{2020}{\natexlab{a}}).

\bibitem[{\citenamefont{An et~al.}(2024)\citenamefont{An, Zhang, Lu, Zhong, and
  Zhang}}]{PLB2024SSJLAn}
\bibinfo{author}{\bibfnamefont{J.~L.} \bibnamefont{An}},
  \bibinfo{author}{\bibfnamefont{K.-Y.} \bibnamefont{Zhang}},
  \bibinfo{author}{\bibfnamefont{Q.}~\bibnamefont{Lu}},
  \bibinfo{author}{\bibfnamefont{S.-Y.} \bibnamefont{Zhong}}, \bibnamefont{and}
  \bibinfo{author}{\bibfnamefont{S.-S.} \bibnamefont{Zhang}},
  \bibinfo{journal}{Phys. Lett. B} \textbf{\bibinfo{volume}{849}},
  \bibinfo{pages}{138422} (\bibinfo{year}{2024}).

\bibitem[{\citenamefont{Zhang et~al.}(2024{\natexlab{a}})\citenamefont{Zhang,
  Huang, Sun, Peng, and Zhang}}]{CPC2024Zhang_48_104105}
\bibinfo{author}{\bibfnamefont{W.}~\bibnamefont{Zhang}},
  \bibinfo{author}{\bibfnamefont{J.~K.} \bibnamefont{Huang}},
  \bibinfo{author}{\bibfnamefont{T.-T.} \bibnamefont{Sun}},
  \bibinfo{author}{\bibfnamefont{J.}~\bibnamefont{Peng}}, \bibnamefont{and}
  \bibinfo{author}{\bibfnamefont{S.~Q.} \bibnamefont{Zhang}},
  \bibinfo{journal}{Chin. Phys. C} \textbf{\bibinfo{volume}{48}},
  \bibinfo{pages}{104105} (\bibinfo{year}{2024}{\natexlab{a}}).

\bibitem[{\citenamefont{M\"oller et~al.}(2009)\citenamefont{M\"oller, Sierk,
  Ichikawa et~al.}}]{PRC2009Moller_79_064304}
\bibinfo{author}{\bibfnamefont{P.}~\bibnamefont{M\"oller}},
  \bibinfo{author}{\bibfnamefont{A.~J.} \bibnamefont{Sierk}},
  \bibinfo{author}{\bibfnamefont{T.}~\bibnamefont{Ichikawa}},
  \bibnamefont{et~al.}, \bibinfo{journal}{Phys. Rev. C}
  \textbf{\bibinfo{volume}{79}}, \bibinfo{pages}{064304}
  (\bibinfo{year}{2009}).

\bibitem[{\citenamefont{Zhou}(2016)}]{PS2016Zhou_91_063008}
\bibinfo{author}{\bibfnamefont{S.~G.} \bibnamefont{Zhou}},
  \bibinfo{journal}{Phys. Scr.} \textbf{\bibinfo{volume}{91}},
  \bibinfo{pages}{063008} (\bibinfo{year}{2016}).

\bibitem[{\citenamefont{Heyde and Wood}(2011)}]{RevModPhys2011}
\bibinfo{author}{\bibfnamefont{K.}~\bibnamefont{Heyde}} \bibnamefont{and}
  \bibinfo{author}{\bibfnamefont{J.~L.} \bibnamefont{Wood}},
  \bibinfo{journal}{Rev. Mod. Phys.} \textbf{\bibinfo{volume}{83}},
  \bibinfo{pages}{1467} (\bibinfo{year}{2011}).

\bibitem[{\citenamefont{Chen et~al.}(2021)\citenamefont{Chen, Sun, Li, and
  Sun}}]{SC2021Chen_64_282011}
\bibinfo{author}{\bibfnamefont{C.}~\bibnamefont{Chen}},
  \bibinfo{author}{\bibfnamefont{Q.~K.} \bibnamefont{Sun}},
  \bibinfo{author}{\bibfnamefont{Y.~X.} \bibnamefont{Li}}, \bibnamefont{and}
  \bibinfo{author}{\bibfnamefont{T.~T.} \bibnamefont{Sun}},
  \bibinfo{journal}{Sci. China-Phys. Mech. Astron.}
  \textbf{\bibinfo{volume}{64}}, \bibinfo{pages}{282011}
  (\bibinfo{year}{2021}).

\bibitem[{\citenamefont{Sun et~al.}(2022{\natexlab{a}})\citenamefont{Sun, Sun,
  Zhang, Zhang, and Chen}}]{CPC2022Sun_46_074106}
\bibinfo{author}{\bibfnamefont{Q.~K.} \bibnamefont{Sun}},
  \bibinfo{author}{\bibfnamefont{T.~T.} \bibnamefont{Sun}},
  \bibinfo{author}{\bibfnamefont{W.}~\bibnamefont{Zhang}},
  \bibinfo{author}{\bibfnamefont{S.~S.} \bibnamefont{Zhang}}, \bibnamefont{and}
  \bibinfo{author}{\bibfnamefont{C.}~\bibnamefont{Chen}},
  \bibinfo{journal}{Chin. Phys. C} \textbf{\bibinfo{volume}{46}},
  \bibinfo{pages}{074106} (\bibinfo{year}{2022}{\natexlab{a}}).

\bibitem[{\citenamefont{Cheng et~al.}(2021)\citenamefont{Cheng, Ge, Cao, and
  Zhang}}]{JPG2021Cheng_48_095106}
\bibinfo{author}{\bibfnamefont{S.~H.} \bibnamefont{Cheng}},
  \bibinfo{author}{\bibfnamefont{Z.~S.} \bibnamefont{Ge}},
  \bibinfo{author}{\bibfnamefont{L.~G.} \bibnamefont{Cao}}, \bibnamefont{and}
  \bibinfo{author}{\bibfnamefont{F.~S.} \bibnamefont{Zhang}},
  \bibinfo{journal}{J. Phys. G} \textbf{\bibinfo{volume}{48}},
  \bibinfo{pages}{095106} (\bibinfo{year}{2021}).

\bibitem[{\citenamefont{Tsunoda et~al.}(2020)\citenamefont{Tsunoda, Otsuka,
  Takayanagi et~al.}}]{Nature2020Tsunoda_587_66}
\bibinfo{author}{\bibfnamefont{N.}~\bibnamefont{Tsunoda}},
  \bibinfo{author}{\bibfnamefont{T.}~\bibnamefont{Otsuka}},
  \bibinfo{author}{\bibfnamefont{K.}~\bibnamefont{Takayanagi}},
  \bibnamefont{et~al.}, \bibinfo{journal}{Nature}
  \textbf{\bibinfo{volume}{587}}, \bibinfo{pages}{66} (\bibinfo{year}{2020}).

\bibitem[{\citenamefont{Li et~al.}(2012{\natexlab{b}})\citenamefont{Li, Meng,
  Ring, Zhao, and Zhou}}]{PRC2012Li_85_024312}
\bibinfo{author}{\bibfnamefont{L.~L.} \bibnamefont{Li}},
  \bibinfo{author}{\bibfnamefont{J.}~\bibnamefont{Meng}},
  \bibinfo{author}{\bibfnamefont{P.}~\bibnamefont{Ring}},
  \bibinfo{author}{\bibfnamefont{E.~G.} \bibnamefont{Zhao}}, \bibnamefont{and}
  \bibinfo{author}{\bibfnamefont{S.~G.} \bibnamefont{Zhou}},
  \bibinfo{journal}{Phys. Rev. C} \textbf{\bibinfo{volume}{85}},
  \bibinfo{pages}{024312} (\bibinfo{year}{2012}{\natexlab{b}}).

\bibitem[{\citenamefont{Zhang et~al.}(2020)\citenamefont{Zhang, Cheoun, Choi,
  and et~al. (DRHBc Mass Table~Collaboration)}}]{PRC2020Zhang_102_024314}
\bibinfo{author}{\bibfnamefont{K.~Y.} \bibnamefont{Zhang}},
  \bibinfo{author}{\bibfnamefont{M.~K.} \bibnamefont{Cheoun}},
  \bibinfo{author}{\bibfnamefont{Y.~B.} \bibnamefont{Choi}}, \bibnamefont{and}
  \bibinfo{author}{\bibnamefont{et~al. (DRHBc Mass Table~Collaboration)}},
  \bibinfo{journal}{Phys. Rev. C} \textbf{\bibinfo{volume}{102}},
  \bibinfo{pages}{024314} (\bibinfo{year}{2020}).

\bibitem[{\citenamefont{Sun et~al.}(2020{\natexlab{b}})\citenamefont{Sun, Zhao,
  and Zhou}}]{NPA2020Sun_1003_122011}
\bibinfo{author}{\bibfnamefont{X.~X.} \bibnamefont{Sun}},
  \bibinfo{author}{\bibfnamefont{J.}~\bibnamefont{Zhao}}, \bibnamefont{and}
  \bibinfo{author}{\bibfnamefont{S.~G.} \bibnamefont{Zhou}},
  \bibinfo{journal}{Nucl. Phys. A} \textbf{\bibinfo{volume}{1003}},
  \bibinfo{pages}{122011} (\bibinfo{year}{2020}{\natexlab{b}}).

\bibitem[{\citenamefont{Sun}(2021)}]{PRC2021Sun_103_054315}
\bibinfo{author}{\bibfnamefont{X.~X.} \bibnamefont{Sun}},
  \bibinfo{journal}{Phys. Rev. C} \textbf{\bibinfo{volume}{103}},
  \bibinfo{pages}{054315} (\bibinfo{year}{2021}).

\bibitem[{\citenamefont{Zhang et~al.}(2023{\natexlab{a}})\citenamefont{Zhang,
  Yang, An, Zhang, Papakonstantinou, Mun, Kim, and
  Yan}}]{PLB2023Zhang_844_138112}
\bibinfo{author}{\bibfnamefont{K.~Y.} \bibnamefont{Zhang}},
  \bibinfo{author}{\bibfnamefont{S.~Q.} \bibnamefont{Yang}},
  \bibinfo{author}{\bibfnamefont{J.~L.} \bibnamefont{An}},
  \bibinfo{author}{\bibfnamefont{S.~S.} \bibnamefont{Zhang}},
  \bibinfo{author}{\bibfnamefont{P.}~\bibnamefont{Papakonstantinou}},
  \bibinfo{author}{\bibfnamefont{M.~H.} \bibnamefont{Mun}},
  \bibinfo{author}{\bibfnamefont{Y.}~\bibnamefont{Kim}}, \bibnamefont{and}
  \bibinfo{author}{\bibfnamefont{H.}~\bibnamefont{Yan}},
  \bibinfo{journal}{Phys. Lett. B} \textbf{\bibinfo{volume}{844}},
  \bibinfo{pages}{138112} (\bibinfo{year}{2023}{\natexlab{a}}).

\bibitem[{\citenamefont{Pan et~al.}(2024)\citenamefont{Pan, Zhang, and
  Zhang}}]{PLB2024Pan}
\bibinfo{author}{\bibfnamefont{C.}~\bibnamefont{Pan}},
  \bibinfo{author}{\bibfnamefont{K.}~\bibnamefont{Zhang}}, \bibnamefont{and}
  \bibinfo{author}{\bibfnamefont{S.}~\bibnamefont{Zhang}},
  \bibinfo{journal}{Phys. Lett. B} \textbf{\bibinfo{volume}{855}},
  \bibinfo{pages}{138792} (\bibinfo{year}{2024}).

\bibitem[{\citenamefont{Sun et~al.}(2018)\citenamefont{Sun, Zhao, and
  Zhou}}]{PLB2018Sun_785_530}
\bibinfo{author}{\bibfnamefont{X.~X.} \bibnamefont{Sun}},
  \bibinfo{author}{\bibfnamefont{J.}~\bibnamefont{Zhao}}, \bibnamefont{and}
  \bibinfo{author}{\bibfnamefont{S.~G.} \bibnamefont{Zhou}},
  \bibinfo{journal}{Phys. Lett. B} \textbf{\bibinfo{volume}{785}},
  \bibinfo{pages}{530} (\bibinfo{year}{2018}).

\bibitem[{\citenamefont{Zhang et~al.}(2019)\citenamefont{Zhang, Wang, and
  Zhang}}]{PRC2019Zhang_100_034312}
\bibinfo{author}{\bibfnamefont{K.~Y.} \bibnamefont{Zhang}},
  \bibinfo{author}{\bibfnamefont{D.~Y.} \bibnamefont{Wang}}, \bibnamefont{and}
  \bibinfo{author}{\bibfnamefont{S.~Q.} \bibnamefont{Zhang}},
  \bibinfo{journal}{Phys. Rev. C} \textbf{\bibinfo{volume}{100}},
  \bibinfo{pages}{034312} (\bibinfo{year}{2019}).

\bibitem[{\citenamefont{In et~al.}(2021)\citenamefont{In, Papakonstantinou,
  Kim, and Hong}}]{IJMPE2021In_30_2150009}
\bibinfo{author}{\bibfnamefont{E.~J.} \bibnamefont{In}},
  \bibinfo{author}{\bibfnamefont{P.}~\bibnamefont{Papakonstantinou}},
  \bibinfo{author}{\bibfnamefont{Y.}~\bibnamefont{Kim}}, \bibnamefont{and}
  \bibinfo{author}{\bibfnamefont{S.~W.} \bibnamefont{Hong}},
  \bibinfo{journal}{Int. J. Mod. Phys. E} \textbf{\bibinfo{volume}{30}},
  \bibinfo{pages}{2150009} (\bibinfo{year}{2021}).

\bibitem[{\citenamefont{Kim et~al.}(2022)\citenamefont{Kim, Mun, Cheoun, and
  Ha}}]{PRC2022Kim_105_034340}
\bibinfo{author}{\bibfnamefont{S.}~\bibnamefont{Kim}},
  \bibinfo{author}{\bibfnamefont{M.~H.} \bibnamefont{Mun}},
  \bibinfo{author}{\bibfnamefont{M.~K.} \bibnamefont{Cheoun}},
  \bibnamefont{and} \bibinfo{author}{\bibfnamefont{E.}~\bibnamefont{Ha}},
  \bibinfo{journal}{Phys. Rev. C} \textbf{\bibinfo{volume}{105}},
  \bibinfo{pages}{034340} (\bibinfo{year}{2022}).

\bibitem[{\citenamefont{Guo et~al.}(2023)\citenamefont{Guo, Pan, Zhao, Du, and
  Zhang}}]{PRC2023Guo_108_014319}
\bibinfo{author}{\bibfnamefont{P.}~\bibnamefont{Guo}},
  \bibinfo{author}{\bibfnamefont{C.}~\bibnamefont{Pan}},
  \bibinfo{author}{\bibfnamefont{Y.~C.} \bibnamefont{Zhao}},
  \bibinfo{author}{\bibfnamefont{X.~K.} \bibnamefont{Du}}, \bibnamefont{and}
  \bibinfo{author}{\bibfnamefont{S.~Q.} \bibnamefont{Zhang}},
  \bibinfo{journal}{Phys. Rev. C} \textbf{\bibinfo{volume}{108}},
  \bibinfo{pages}{014319} (\bibinfo{year}{2023}).

\bibitem[{\citenamefont{Zhang et~al.}(2023{\natexlab{b}})\citenamefont{Zhang,
  Niu, Sun, and Xia}}]{PRC2023Zhang_108_024310}
\bibinfo{author}{\bibfnamefont{X.~Y.} \bibnamefont{Zhang}},
  \bibinfo{author}{\bibfnamefont{Z.~M.} \bibnamefont{Niu}},
  \bibinfo{author}{\bibfnamefont{W.}~\bibnamefont{Sun}}, \bibnamefont{and}
  \bibinfo{author}{\bibfnamefont{X.~W.} \bibnamefont{Xia}},
  \bibinfo{journal}{Phys. Rev. C} \textbf{\bibinfo{volume}{108}},
  \bibinfo{pages}{024310} (\bibinfo{year}{2023}{\natexlab{b}}).

\bibitem[{\citenamefont{Mun et~al.}(2023)\citenamefont{Mun, Kim, Cheoun, So,
  Choi, and Ha}}]{PLB2023Mun_847_138298}
\bibinfo{author}{\bibfnamefont{M.~H.} \bibnamefont{Mun}},
  \bibinfo{author}{\bibfnamefont{S.}~\bibnamefont{Kim}},
  \bibinfo{author}{\bibfnamefont{M.~K.} \bibnamefont{Cheoun}},
  \bibinfo{author}{\bibfnamefont{W.~Y.} \bibnamefont{So}},
  \bibinfo{author}{\bibfnamefont{S.}~\bibnamefont{Choi}}, \bibnamefont{and}
  \bibinfo{author}{\bibfnamefont{E.}~\bibnamefont{Ha}}, \bibinfo{journal}{Phys.
  Lett. B} \textbf{\bibinfo{volume}{847}}, \bibinfo{pages}{138298}
  (\bibinfo{year}{2023}).

\bibitem[{\citenamefont{Zhang et~al.}(2023{\natexlab{c}})\citenamefont{Zhang,
  Papakonstantinou, Mun, Kim, Yan, and Sun}}]{PRC2023Zhang_107_L041303}
\bibinfo{author}{\bibfnamefont{K.~Y.} \bibnamefont{Zhang}},
  \bibinfo{author}{\bibfnamefont{P.}~\bibnamefont{Papakonstantinou}},
  \bibinfo{author}{\bibfnamefont{M.~H.} \bibnamefont{Mun}},
  \bibinfo{author}{\bibfnamefont{Y.}~\bibnamefont{Kim}},
  \bibinfo{author}{\bibfnamefont{H.}~\bibnamefont{Yan}}, \bibnamefont{and}
  \bibinfo{author}{\bibfnamefont{X.~X.} \bibnamefont{Sun}},
  \bibinfo{journal}{Phys. Rev. C} \textbf{\bibinfo{volume}{107}},
  \bibinfo{pages}{L041303} (\bibinfo{year}{2023}{\natexlab{c}}).

\bibitem[{\citenamefont{Zheng et~al.}(2024)\citenamefont{Zheng, Sun, Shen, and
  Geng}}]{CPC2024Zheng_48_014107}
\bibinfo{author}{\bibfnamefont{R.~Y.} \bibnamefont{Zheng}},
  \bibinfo{author}{\bibfnamefont{X.~X.} \bibnamefont{Sun}},
  \bibinfo{author}{\bibfnamefont{G.~F.} \bibnamefont{Shen}}, \bibnamefont{and}
  \bibinfo{author}{\bibfnamefont{L.~S.} \bibnamefont{Geng}},
  \bibinfo{journal}{Chin. Phys. C} \textbf{\bibinfo{volume}{48}},
  \bibinfo{pages}{014107} (\bibinfo{year}{2024}).

\bibitem[{\citenamefont{Zhang et~al.}(2021)\citenamefont{Zhang, He, Meng, Pan,
  Shen, Wang, and Zhang}}]{PRC2021Zhang_104_l021301}
\bibinfo{author}{\bibfnamefont{K.~Y.} \bibnamefont{Zhang}},
  \bibinfo{author}{\bibfnamefont{X.~T.} \bibnamefont{He}},
  \bibinfo{author}{\bibfnamefont{J.}~\bibnamefont{Meng}},
  \bibinfo{author}{\bibfnamefont{C.}~\bibnamefont{Pan}},
  \bibinfo{author}{\bibfnamefont{C.~W.} \bibnamefont{Shen}},
  \bibinfo{author}{\bibfnamefont{C.}~\bibnamefont{Wang}}, \bibnamefont{and}
  \bibinfo{author}{\bibfnamefont{S.~Q.} \bibnamefont{Zhang}},
  \bibinfo{journal}{Phys. Rev. C} \textbf{\bibinfo{volume}{104}},
  \bibinfo{pages}{l021301} (\bibinfo{year}{2021}).

\bibitem[{\citenamefont{Pan et~al.}(2021)\citenamefont{Pan, Zhang, Chong
  et~al.}}]{PRC2021Pan_104_024331}
\bibinfo{author}{\bibfnamefont{C.}~\bibnamefont{Pan}},
  \bibinfo{author}{\bibfnamefont{K.~Y.} \bibnamefont{Zhang}},
  \bibinfo{author}{\bibfnamefont{P.~S.} \bibnamefont{Chong}},
  \bibnamefont{et~al.}, \bibinfo{journal}{Phys. Rev. C}
  \textbf{\bibinfo{volume}{104}}, \bibinfo{pages}{024331}
  (\bibinfo{year}{2021}).

\bibitem[{\citenamefont{He et~al.}(2021)\citenamefont{He, Wang, Zhang, and
  Shen}}]{CPC2021He_45_101001}
\bibinfo{author}{\bibfnamefont{X.~T.} \bibnamefont{He}},
  \bibinfo{author}{\bibfnamefont{C.}~\bibnamefont{Wang}},
  \bibinfo{author}{\bibfnamefont{K.~Y.} \bibnamefont{Zhang}}, \bibnamefont{and}
  \bibinfo{author}{\bibfnamefont{C.~W.} \bibnamefont{Shen}},
  \bibinfo{journal}{Chin. Phys. C} \textbf{\bibinfo{volume}{45}},
  \bibinfo{pages}{101001} (\bibinfo{year}{2021}).

\bibitem[{\citenamefont{Xiao et~al.}(2023)\citenamefont{Xiao, Xu, Zheng, Sun,
  Geng, and Zhang}}]{PLB2023Xiao_845_138160}
\bibinfo{author}{\bibfnamefont{Y.}~\bibnamefont{Xiao}},
  \bibinfo{author}{\bibfnamefont{S.~Z.} \bibnamefont{Xu}},
  \bibinfo{author}{\bibfnamefont{R.~Y.} \bibnamefont{Zheng}},
  \bibinfo{author}{\bibfnamefont{X.~X.} \bibnamefont{Sun}},
  \bibinfo{author}{\bibfnamefont{L.~S.} \bibnamefont{Geng}}, \bibnamefont{and}
  \bibinfo{author}{\bibfnamefont{S.~S.} \bibnamefont{Zhang}},
  \bibinfo{journal}{Phys. Lett. B} \textbf{\bibinfo{volume}{845}},
  \bibinfo{pages}{138160} (\bibinfo{year}{2023}).

\bibitem[{\citenamefont{Lu et~al.}(2024)\citenamefont{Lu, Zhang, and
  Zhang}}]{PLB2024SSZhang}
\bibinfo{author}{\bibfnamefont{Q.}~\bibnamefont{Lu}},
  \bibinfo{author}{\bibfnamefont{K.-Y.} \bibnamefont{Zhang}}, \bibnamefont{and}
  \bibinfo{author}{\bibfnamefont{S.-S.} \bibnamefont{Zhang}},
  \bibinfo{journal}{Phys. Lett. B} \textbf{\bibinfo{volume}{856}},
  \bibinfo{pages}{138922} (\bibinfo{year}{2024}).

\bibitem[{\citenamefont{Pan et~al.}(2019)\citenamefont{Pan, Zhang, and
  Zhang}}]{IJMPE2019Pan_28_1950082}
\bibinfo{author}{\bibfnamefont{C.}~\bibnamefont{Pan}},
  \bibinfo{author}{\bibfnamefont{K.~Y.} \bibnamefont{Zhang}}, \bibnamefont{and}
  \bibinfo{author}{\bibfnamefont{S.~Q.} \bibnamefont{Zhang}},
  \bibinfo{journal}{Int. J. Mod. Phys. E} \textbf{\bibinfo{volume}{28}},
  \bibinfo{pages}{1950082} (\bibinfo{year}{2019}).

\bibitem[{\citenamefont{Zhang et~al.}(2022{\natexlab{b}})\citenamefont{Zhang,
  Pan, and Zhang}}]{PRC2022Zhang_106_024302}
\bibinfo{author}{\bibfnamefont{K.~Y.} \bibnamefont{Zhang}},
  \bibinfo{author}{\bibfnamefont{C.}~\bibnamefont{Pan}}, \bibnamefont{and}
  \bibinfo{author}{\bibfnamefont{S.~Q.} \bibnamefont{Zhang}},
  \bibinfo{journal}{Phys. Rev. C} \textbf{\bibinfo{volume}{106}},
  \bibinfo{pages}{024302} (\bibinfo{year}{2022}{\natexlab{b}}).

\bibitem[{\citenamefont{Sun and Zhou}(2021)}]{SB2021Sun_66_2072}
\bibinfo{author}{\bibfnamefont{X.~X.} \bibnamefont{Sun}} \bibnamefont{and}
  \bibinfo{author}{\bibfnamefont{S.~G.} \bibnamefont{Zhou}},
  \bibinfo{journal}{Sci. Bull.} \textbf{\bibinfo{volume}{66}},
  \bibinfo{pages}{2072} (\bibinfo{year}{2021}).

\bibitem[{\citenamefont{Sun et~al.}(2022{\natexlab{b}})\citenamefont{Sun,
  Zhang, Pan, Fan, Zhang, and Li}}]{CPC2022Sun_46_064103}
\bibinfo{author}{\bibfnamefont{W.}~\bibnamefont{Sun}},
  \bibinfo{author}{\bibfnamefont{K.~Y.} \bibnamefont{Zhang}},
  \bibinfo{author}{\bibfnamefont{C.}~\bibnamefont{Pan}},
  \bibinfo{author}{\bibfnamefont{X.~H.} \bibnamefont{Fan}},
  \bibinfo{author}{\bibfnamefont{S.~Q.} \bibnamefont{Zhang}}, \bibnamefont{and}
  \bibinfo{author}{\bibfnamefont{Z.~P.} \bibnamefont{Li}},
  \bibinfo{journal}{Chin. Phys. C} \textbf{\bibinfo{volume}{46}},
  \bibinfo{pages}{064103} (\bibinfo{year}{2022}{\natexlab{b}}).

\bibitem[{\citenamefont{Khuyagbaatar et~al.}(2020)\citenamefont{Khuyagbaatar,
  Yakushev, D\"ullmann, Ackermann, Andersson, Asai, Block, Boll, Brand, Cox
  et~al.}}]{PRC2020Khuyagbaatar_102_064602}
\bibinfo{author}{\bibfnamefont{J.}~\bibnamefont{Khuyagbaatar}},
  \bibinfo{author}{\bibfnamefont{A.}~\bibnamefont{Yakushev}},
  \bibinfo{author}{\bibfnamefont{C.~E.} \bibnamefont{D\"ullmann}},
  \bibinfo{author}{\bibfnamefont{D.}~\bibnamefont{Ackermann}},
  \bibinfo{author}{\bibfnamefont{L.-L.} \bibnamefont{Andersson}},
  \bibinfo{author}{\bibfnamefont{M.}~\bibnamefont{Asai}},
  \bibinfo{author}{\bibfnamefont{M.}~\bibnamefont{Block}},
  \bibinfo{author}{\bibfnamefont{R.~A.} \bibnamefont{Boll}},
  \bibinfo{author}{\bibfnamefont{H.}~\bibnamefont{Brand}},
  \bibinfo{author}{\bibfnamefont{D.~M.} \bibnamefont{Cox}},
  \bibnamefont{et~al.}, \bibinfo{journal}{Phys. Rev. C}
  \textbf{\bibinfo{volume}{102}}, \bibinfo{pages}{064602}
  (\bibinfo{year}{2020}).

\bibitem[{\citenamefont{Li et~al.}(2012{\natexlab{c}})\citenamefont{Li, Meng,
  Ring, Zhao, and Zhou}}]{CPL2012Li_29_042101}
\bibinfo{author}{\bibfnamefont{L.~L.} \bibnamefont{Li}},
  \bibinfo{author}{\bibfnamefont{J.}~\bibnamefont{Meng}},
  \bibinfo{author}{\bibfnamefont{P.}~\bibnamefont{Ring}},
  \bibinfo{author}{\bibfnamefont{E.~G.} \bibnamefont{Zhao}}, \bibnamefont{and}
  \bibinfo{author}{\bibfnamefont{S.~G.} \bibnamefont{Zhou}},
  \bibinfo{journal}{Chin. Phys. Lett.} \textbf{\bibinfo{volume}{29}},
  \bibinfo{pages}{042101} (\bibinfo{year}{2012}{\natexlab{c}}).

\bibitem[{\citenamefont{Price and Walker}(1987)}]{PRC1987Price_36_354}
\bibinfo{author}{\bibfnamefont{C.~E.} \bibnamefont{Price}} \bibnamefont{and}
  \bibinfo{author}{\bibfnamefont{G.~E.} \bibnamefont{Walker}},
  \bibinfo{journal}{Phys. Rev. C} \textbf{\bibinfo{volume}{36}},
  \bibinfo{pages}{354} (\bibinfo{year}{1987}).

\bibitem[{\citenamefont{Perez-Martin and
  Robledo}(2008)}]{PRC2008Perez-Martin_78_014304}
\bibinfo{author}{\bibfnamefont{S.}~\bibnamefont{Perez-Martin}}
  \bibnamefont{and} \bibinfo{author}{\bibfnamefont{L.~M.}
  \bibnamefont{Robledo}}, \bibinfo{journal}{Phys. Rev. C}
  \textbf{\bibinfo{volume}{78}}, \bibinfo{pages}{014304}
  (\bibinfo{year}{2008}).

\bibitem[{\citenamefont{Pan et~al.}(2022)\citenamefont{Pan, Cheoun, Choi, and
  et~al. (DRHBc Mass Table~Collaboration)}}]{PRC2022Pan_106_014316}
\bibinfo{author}{\bibfnamefont{C.}~\bibnamefont{Pan}},
  \bibinfo{author}{\bibfnamefont{M.~K.} \bibnamefont{Cheoun}},
  \bibinfo{author}{\bibfnamefont{Y.~B.} \bibnamefont{Choi}}, \bibnamefont{and}
  \bibinfo{author}{\bibnamefont{et~al. (DRHBc Mass Table~Collaboration)}},
  \bibinfo{journal}{Phys. Rev. C} \textbf{\bibinfo{volume}{106}},
  \bibinfo{pages}{014316} (\bibinfo{year}{2022}).

\bibitem[{\citenamefont{Zhou et~al.}(2003)\citenamefont{Zhou, Meng, and
  Ring}}]{PRC2003Zhou_68_034323}
\bibinfo{author}{\bibfnamefont{S.~G.} \bibnamefont{Zhou}},
  \bibinfo{author}{\bibfnamefont{J.}~\bibnamefont{Meng}}, \bibnamefont{and}
  \bibinfo{author}{\bibfnamefont{P.}~\bibnamefont{Ring}},
  \bibinfo{journal}{Phys. Rev. C} \textbf{\bibinfo{volume}{68}},
  \bibinfo{pages}{034323} (\bibinfo{year}{2003}).

\bibitem[{\citenamefont{Wang et~al.}(2022)\citenamefont{Wang, Sun, and
  Zhou}}]{CPC2022Wang_46_024107}
\bibinfo{author}{\bibfnamefont{X.~Q.} \bibnamefont{Wang}},
  \bibinfo{author}{\bibfnamefont{X.~X.} \bibnamefont{Sun}}, \bibnamefont{and}
  \bibinfo{author}{\bibfnamefont{S.~G.} \bibnamefont{Zhou}},
  \bibinfo{journal}{Chin. Phys. C} \textbf{\bibinfo{volume}{46}},
  \bibinfo{pages}{024107} (\bibinfo{year}{2022}).

\bibitem[{\citenamefont{Nrtershäuser et~al.}(2009)\citenamefont{Nrtershäuser,
  Tiedemann, Žáková et~al.}}]{PRL2009Nortershauser_102_062503}
\bibinfo{author}{\bibfnamefont{W.}~\bibnamefont{Nrtershäuser}},
  \bibinfo{author}{\bibfnamefont{D.}~\bibnamefont{Tiedemann}},
  \bibinfo{author}{\bibfnamefont{M.}~\bibnamefont{Žáková}},
  \bibnamefont{et~al.}, \bibinfo{journal}{Phys. Rev. Lett.}
  \textbf{\bibinfo{volume}{102}}, \bibinfo{pages}{062503}
  (\bibinfo{year}{2009}).

\bibitem[{\citenamefont{Zhang et~al.}(2022{\natexlab{c}})\citenamefont{Zhang,
  Cheoun, Choi, and et~al. (DRHBc Mass
  Table~Collaboration)}}]{ADNDT2022Zhang_144_101488}
\bibinfo{author}{\bibfnamefont{K.~Y.} \bibnamefont{Zhang}},
  \bibinfo{author}{\bibfnamefont{M.~K.} \bibnamefont{Cheoun}},
  \bibinfo{author}{\bibfnamefont{Y.~B.} \bibnamefont{Choi}}, \bibnamefont{and}
  \bibinfo{author}{\bibnamefont{et~al. (DRHBc Mass Table~Collaboration)}},
  \bibinfo{journal}{At. Data Nucl. Data Tables} \textbf{\bibinfo{volume}{144}},
  \bibinfo{pages}{101488} (\bibinfo{year}{2022}{\natexlab{c}}).

\bibitem[{\citenamefont{Guo et~al.}(2024)\citenamefont{Guo, Cao, Chen, and
  et~al. (DRHBc Mass Table~Collaboration)}}]{ADNDT2024Guo}
\bibinfo{author}{\bibfnamefont{P.}~\bibnamefont{Guo}},
  \bibinfo{author}{\bibfnamefont{X.}~\bibnamefont{Cao}},
  \bibinfo{author}{\bibfnamefont{K.}~\bibnamefont{Chen}}, \bibnamefont{and}
  \bibinfo{author}{\bibnamefont{et~al. (DRHBc Mass Table~Collaboration)}},
  \bibinfo{journal}{At. Data Nucl. Data Tables} \textbf{\bibinfo{volume}{158}},
  \bibinfo{pages}{101661} (\bibinfo{year}{2024}).

\bibitem[{\citenamefont{Wang et~al.}(2021)\citenamefont{Wang, Huang, Kondev,
  Audi, and Naimi}}]{CPC2021Wang_45_030003}
\bibinfo{author}{\bibfnamefont{M.}~\bibnamefont{Wang}},
  \bibinfo{author}{\bibfnamefont{W.~J.} \bibnamefont{Huang}},
  \bibinfo{author}{\bibfnamefont{F.~G.} \bibnamefont{Kondev}},
  \bibinfo{author}{\bibfnamefont{G.}~\bibnamefont{Audi}}, \bibnamefont{and}
  \bibinfo{author}{\bibfnamefont{S.}~\bibnamefont{Naimi}},
  \bibinfo{journal}{Chin. Phys. C} \textbf{\bibinfo{volume}{45}},
  \bibinfo{pages}{030003} (\bibinfo{year}{2021}).

\bibitem[{\citenamefont{Xia et~al.}(2018)\citenamefont{Xia, Lim, Zhao
  et~al.}}]{ADNDT2018Xia_121_1}
\bibinfo{author}{\bibfnamefont{X.~W.} \bibnamefont{Xia}},
  \bibinfo{author}{\bibfnamefont{Y.}~\bibnamefont{Lim}},
  \bibinfo{author}{\bibfnamefont{P.~W.} \bibnamefont{Zhao}},
  \bibnamefont{et~al.}, \bibinfo{journal}{At. Data Nucl. Data Tables}
  \textbf{\bibinfo{volume}{121}}, \bibinfo{pages}{1} (\bibinfo{year}{2018}).

\bibitem[{\citenamefont{Zhang et~al.}(2025{\natexlab{b}})\citenamefont{Wu,
  Zhang, Peng, and Huang}}]{Particle2026WZhang}
\bibinfo{author}{\bibfnamefont{L.} \bibnamefont{Wu}},
  \bibinfo{author}{\bibfnamefont{W.} \bibnamefont{Zhang}},
  \bibinfo{author}{\bibfnamefont{J.} \bibnamefont{Peng}}, \bibnamefont{and}
  \bibinfo{author}{\bibfnamefont{J.} \bibnamefont{Huang}},
  \bibinfo{journal}{Particles} \textbf{\bibinfo{volume}{8}},
  \bibinfo{pages}{19} (\bibinfo{year}{2025}{\natexlab{b}}).

\bibitem[{\citenamefont{Zhang et~al.}(2024{\natexlab{b}})\citenamefont{Zhang,
  Liu, Zhang, and Yao}}]{PRC2024JMYao}
\bibinfo{author}{\bibfnamefont{Y.~X.} \bibnamefont{Zhang}},
  \bibinfo{author}{\bibfnamefont{B.~R.} \bibnamefont{Liu}},
  \bibinfo{author}{\bibfnamefont{K.~Y.} \bibnamefont{Zhang}}, \bibnamefont{and}
  \bibinfo{author}{\bibfnamefont{J.~M.} \bibnamefont{Yao}},
  \bibinfo{journal}{Phys. Rev. C} \textbf{\bibinfo{volume}{110}},
  \bibinfo{pages}{024302} (\bibinfo{year}{2024}{\natexlab{b}}).

\bibitem[{\citenamefont{Casten}(2000)}]{Casten2000nuclear}
\bibinfo{author}{\bibfnamefont{R.}~\bibnamefont{Casten}},
  \emph{\bibinfo{title}{Nuclear structure from a simple perspective}},
  vol.~\bibinfo{volume}{23} (\bibinfo{publisher}{Oxford university press},
  \bibinfo{year}{2000}).

\bibitem[{\citenamefont{Sugawara}(2022)}]{PRC2022Sugawara_106_024301}
\bibinfo{author}{\bibfnamefont{M.}~\bibnamefont{Sugawara}},
  \bibinfo{journal}{Phys. Rev. C} \textbf{\bibinfo{volume}{106}},
  \bibinfo{pages}{024301} (\bibinfo{year}{2022}).

\bibitem[{\citenamefont{Zhang et~al.}(2023{\natexlab{d}})\citenamefont{Zhang,
  Zhang, and Meng}}]{PRC2023KYZhang_TRHB}
\bibinfo{author}{\bibfnamefont{K.~Y.} \bibnamefont{Zhang}},
  \bibinfo{author}{\bibfnamefont{S.~Q.} \bibnamefont{Zhang}}, \bibnamefont{and}
  \bibinfo{author}{\bibfnamefont{J.}~\bibnamefont{Meng}},
  \bibinfo{journal}{Phys. Rev. C} \textbf{\bibinfo{volume}{108}},
  \bibinfo{pages}{L041301} (\bibinfo{year}{2023}{\natexlab{d}}).

\bibitem[{\citenamefont{Geithner et~al.}(2008)\citenamefont{Geithner, Neff,
  Audi, Blaum, Delahaye, Feldmeier, George, Guénaut, Herfurth, Herlert
  et~al.}}]{PRL2008Geithner_101_252502}
\bibinfo{author}{\bibfnamefont{W.}~\bibnamefont{Geithner}},
  \bibinfo{author}{\bibfnamefont{T.}~\bibnamefont{Neff}},
  \bibinfo{author}{\bibfnamefont{G.}~\bibnamefont{Audi}},
  \bibinfo{author}{\bibfnamefont{K.}~\bibnamefont{Blaum}},
  \bibinfo{author}{\bibfnamefont{P.}~\bibnamefont{Delahaye}},
  \bibinfo{author}{\bibfnamefont{H.}~\bibnamefont{Feldmeier}},
  \bibinfo{author}{\bibfnamefont{S.}~\bibnamefont{George}},
  \bibinfo{author}{\bibfnamefont{C.}~\bibnamefont{Guénaut}},
  \bibinfo{author}{\bibfnamefont{F.}~\bibnamefont{Herfurth}},
  \bibinfo{author}{\bibfnamefont{A.}~\bibnamefont{Herlert}},
  \bibnamefont{et~al.}, \bibinfo{journal}{Phys. Rev. Lett.}
  \textbf{\bibinfo{volume}{101}} (\bibinfo{year}{2008}).

\bibitem[{\citenamefont{Angeli and Marinova}(2013)}]{ADNDT2013Angeli_99_69}
\bibinfo{author}{\bibfnamefont{I.}~\bibnamefont{Angeli}} \bibnamefont{and}
  \bibinfo{author}{\bibfnamefont{K.}~\bibnamefont{Marinova}},
  \bibinfo{journal}{At. Data Nucl. Data Tables} \textbf{\bibinfo{volume}{99}},
  \bibinfo{pages}{69} (\bibinfo{year}{2013}).

\bibitem[{\citenamefont{Li et~al.}(2021)\citenamefont{Li, Luo, and
  Wang}}]{ADNDT2021Li_140_101440}
\bibinfo{author}{\bibfnamefont{T.}~\bibnamefont{Li}},
  \bibinfo{author}{\bibfnamefont{Y.~N.} \bibnamefont{Luo}}, \bibnamefont{and}
  \bibinfo{author}{\bibfnamefont{N.}~\bibnamefont{Wang}}, \bibinfo{journal}{At.
  Data Nucl. Data Tables} \textbf{\bibinfo{volume}{140}},
  \bibinfo{pages}{101440} (\bibinfo{year}{2021}).

\end{thebibliography}

\end{document}